\documentclass[alpha-refs]{wiley-article}

\usepackage{siunitx}
\usepackage{graphicx, subfigure}
\usepackage{enumitem}
\usepackage{algorithm}
\usepackage{algpseudocode}
\usepackage{enumerate}
\usepackage{float}
\usepackage[toc,page]{appendix}
\usepackage{graphicx,epstopdf}
\usepackage{natbib}
\DeclareMathOperator*{\argmin}{arg\,min}

\title{MCMC methods for inference in a mathematical model of pulmonary circulation}

\author[1\authfn{1}]{L. Mihaela Paun}
\author[2\authfn{2}]{M. Umar Qureshi}
\author[2\authfn{2}]{Mitchel Colebank}
\author[2\authfn{1}]{Nicholas A. Hill}
\author[2\authfn{2}]{Mette S. Olufsen}
\author[2\authfn{2}]{Mansoor A. Haider}
\author[2\authfn{1}]{Dirk Husmeier}

\affil[1]{School of Mathematics and Statistics, University of Glasgow, Glasgow G12 8SQ, UK}
\affil[2]{Department of Mathematics, NC State University, Raleigh, NC 27695, USA}

\corraddress{L. Mihaela Paun, School of Mathematics and Statistics, University of Glasgow, Glasgow G12 8SQ, UK}
\corremail{l.paun.1@research.gla.ac.uk}

\fundinginfo{Engineering and Physical Sciences Research Council (EPSRC) of the UK,  Grant/Award Number: EP/N014642/1; National Science Foundation, Grant/Award Number: NSF-DMS \# 1615820}

\runningauthor{Paun et al.}

\makeatletter
\def\therule{\makebox[\algorithmicindent][l]{\hspace*{.5em}\vrule height .75\baselineskip depth .25\baselineskip}}%

\newtoks\therules
\therules={}
\def\appendto#1#2{\expandafter#1\expandafter{\the#1#2}}
\def\gobblefirst#1{
  #1\expandafter\expandafter\expandafter{\expandafter\@gobble\the#1}}%
\def\LState{\State\unskip\the\therules}
\def\pushindent{\appendto\therules\therule}%
\def\popindent{\gobblefirst\therules}%
\def\printindent{\unskip\the\therules}%
\def\printandpush{\printindent\pushindent}%
\def\popandprint{\popindent\printindent}%

\algdef{SE}[WHILE]{While}{EndWhile}[1]
  {\printandpush\algorithmicwhile\ #1\ \algorithmicdo}
  {\popandprint\algorithmicend\ \algorithmicwhile}%
\algdef{SE}[FOR]{For}{EndFor}[1]
  {\printandpush\algorithmicfor\ #1\ \algorithmicdo}
  {\popandprint\algorithmicend\ \algorithmicfor}%
\algdef{S}[FOR]{ForAll}[1]
  {\printindent\algorithmicforall\ #1\ \algorithmicdo}%
\algdef{SE}[LOOP]{Loop}{EndLoop}
  {\printandpush\algorithmicloop}
  {\popandprint\algorithmicend\ \algorithmicloop}%
\algdef{SE}[REPEAT]{Repeat}{Until}
  {\printandpush\algorithmicrepeat}[1]
  {\popandprint\algorithmicuntil\ #1}%
\algdef{SE}[IF]{If}{EndIf}[1]
  {\printandpush\algorithmicif\ #1\ \algorithmicthen}
  {\popandprint\algorithmicend\ \algorithmicif}%
\algdef{C}[IF]{IF}{ElsIf}[1]
  {\popandprint\pushindent\algorithmicelse\ \algorithmicif\ #1\ \algorithmicthen}%
\algdef{Ce}[ELSE]{IF}{Else}{EndIf}
  {\popandprint\pushindent\algorithmicelse}%
\algdef{SE}[PROCEDURE]{Procedure}{EndProcedure}[2]
   {\printandpush\algorithmicprocedure\ \textproc{#1}\ifthenelse{\equal{#2}{}}{}{(#2)}}%
   {\popandprint\algorithmicend\ \algorithmicprocedure}%
\algdef{SE}[FUNCTION]{Function}{EndFunction}[2]
   {\printandpush\algorithmicfunction\ \textproc{#1}\ifthenelse{\equal{#2}{}}{}{(#2)}}%
   {\popandprint\algorithmicend\ \algorithmicfunction}%
\makeatother

\begin{document}

\maketitle

\begin{abstract}
This study performs parameter inference in a partial differential equations system of pulmonary circulation. We use a fluid dynamics network model that takes selected parameter values and mimics the behaviour of the pulmonary haemodynamics under normal physiological and pathological conditions. This is of medical interest as it enables tracking the progression of pulmonary hypertension. We show how we make the fluids model tractable by reducing the parameter dimension from a 55D to a 5D problem. The Delayed Rejection Adaptive Metropolis (DRAM) algorithm, coupled with constraint nonlinear optimization is successfully used to learn the parameter values and quantify the uncertainty in the parameter estimates. To accommodate for different magnitudes of the parameter values, we introduce an improved parameter scaling technique in the DRAM algorithm. Formal convergence diagnostics are employed to check for convergence of the Markov chains. Additionally, we perform model selection using different information criteria, including Watanabe Akaike Information Criteria.

\end{abstract}

\section{Introduction}
The cardiovascular circulation is composed of the systemic circulation and the pulmonary circulation. Extensive work has been done to model the systemic circulation \citep{ma1}, however pulmonary hypertension is one of the leading causes of right heart failure \citep{ma2}. Our work focuses on predicting the observed haemodynamic behaviour in the pulmonary circulation under normal and pathological conditions (hypoxia). Hypoxia is a pathological condition in which the body tissues are not sufficiently well oxygenated, leading to pulmonary hypertension.

Pulmonary hypertension (PH) is characterised by high mean blood pressure in the lungs (above $25 \textrm{ mmHg}$). PH leads to vascular remodelling, including stiffening, thickening, constriction of the small and large pulmonary arteries and microvascular rarefaction \citep{PulmCirc}. Microvascular rarefaction occurs in patients suffering of pulmonary hypertension, and is a pathological condition in which there are fewer capillaries per unit volume of body tissue.
Developing a reliable predictive model for the pulmonary haemodynamics allows us to assist clinicians in diagnosing and treating pulmonary hypertension in a systematic manner. In addition, it helps reduce the number of invasive procedures for patients as, currently, pulmonary pressure is measured invasively via right heart catheterization \citep{Tabima}. In our problem we test the model in the context of data from mice, however, the work can be extended to human data. 

One of the most important aspects in cardiovascular modelling is parameter inference and uncertainty quantification \citep{Eck}. This is because the number of parameters increases significantly when modelling several interconnected components at different scales. What is more, the limited amount of data available makes parameter estimation challenging in terms of finding physiologically sensible parameters.

Inferring key parameters for disease diagnosis and treatment planning is therefore an essential and challenging step in predicting the observed haemodynamics. One such parameter is the arterial stiffness, which is significantly higher for patients having pulmonary hypertension. Often, the parameters cannot be measured in-vivo, which creates the need for them to be learnt indirectly from the blood flow and pressure measurements. 

In this paper, we present statistical techniques for inference of parameters of a mathematical model in order to reliably simulate and predict blood pressure and flow in the lungs. We first employ a nonlinear constraint optimization scheme to find the maximum likelihood estimates. We then quantify the uncertainty around these estimates by approximately sampling parameters from their posterior distribution using the Delayed Rejection Adaptive Metropolis (DRAM) algorithm and the Adaptive Metropolis (AM) algorithm.

Simulating blood flow and pressure in the 1D fluid-structure model described in the next section, requires solving a system of nonlinear partial differential equations (PDEs), which gets computationally expensive as the number of vessels increases in the model or a component model is made more complex. Therefore, starting the MCMC algorithm from optimised values allows us to save computational time. The goal of this paper is to show how optimization, combined with uncertainty quantification, can be used to learn the parameters relevant for pulmonary disease detection and, thus, to reliably predict the measured blood flow and pressure in the lungs. We show how by making a few restrictive assumptions about the parameters characterising the vessels' geometry or the boundary conditions in the PDEs, we manage to reduce the parameter dimension from a 55D problem to 5D. Furthermore, we illustrate via pseudocode how to deal with different parameter magnitudes by incorporating a parameter scaling technique in the MCMC Matlab toolbox \citep{DRAMtoolbox}. We test this code on real data coming from a healthy and a hypoxic mouse. Furthermore, we show the importance of parameter scaling on a concrete example, and the consequences of not including the scaling in the analysis. 

The remainder of the paper is structured as follows: Section 2 introduces the mathematical model, Section 3 describes the statistical model used in this study, Section 4 provides an overview of the methodology used, including the DRAM algorithm allowing for parameter scaling, Section 5 gives a few details of the computer simulations, Section 6 presents results of the methods applied on real data, and we finalise the paper by a discussion of the main findings of the study, as well as limitations and further work.

\section{Mathematical model}{\label{sssec:mathsmodel}}
This study predicts  pulmonary arterial  flow and pressure using a 1D fluid dynamics network model, described in detail in our recent studies \cite{QureshiMice} and \cite{NewPaper}. The model is derived from the incompressible axisymmetric Navier--Stokes equations for a Newtonian fluid, and coupled with a constitutive wall model predicting elasticity (i.e. stiffness) of the blood vessels. 
In addition, the model assumes that the vessels are cylindrical and tapering, and that the wavelength is significantly longer than vessels' radii. We use a tapering factor for the large arteries, as there is evidence of the vessels radii decreasing along their length and this was not quantified during the segmentation process. Under these assumptions, conservation of mass and momentum give
\begin{equation}{\label{eq:1}}
    \frac{\partial{A}}{\partial{t}} + \frac{\partial{q}}{\partial{x}} = 0, \hspace{0.5in} \frac{\partial{q}}{\partial{t}} + \frac{\partial}{\partial{x}} \frac{q^2}{A} + \frac{A}{\rho}\frac{\partial{p}}{\partial{x}}  = -\frac{2\pi\mu r}{\delta}\frac{q}{A},
\end{equation}
where $x \textrm{ (cm)}$ and $t \textrm{ (s)}$ are the axial and temporal coordinates, $p \textrm{ (mmHg)}$ is the blood pressure, $q \textrm{ (ml/s)}$ is the blood flow rate, $A \textrm{ (cm}^2)$ is the cross-sectional area, $\rho = 1.055 \textrm{ g/ml}$ is the blood density, $\mu = 0.049\, \textrm{g/(cm\,s)}$ is the viscosity and $\delta = \sqrt{\mu T/2\pi\rho} \textrm{ } \textrm{cm}$ is the boundary-layer thickness of velocity profile. 

The arterial walls are assumed to be homogeneous, isotropic and thin. We use a linear wall model, where the relationship between pressure, $p$ (stress) and cross-sectional area $A$ deformation (strain) can be expressed as follows:
\begin{equation}{\label{eq:2}}
   p = p_0 + \frac{4}{3}f \left(1-\sqrt \frac{A_0}{A}\right) 
\end{equation}
where $A_0$ is the unstressed vessel cross-sectional area, and $f \textrm{ (mmHg)}$ is the arterial network stiffness.

\begin{figure}[h!]
\centering
\includegraphics[scale=0.35]{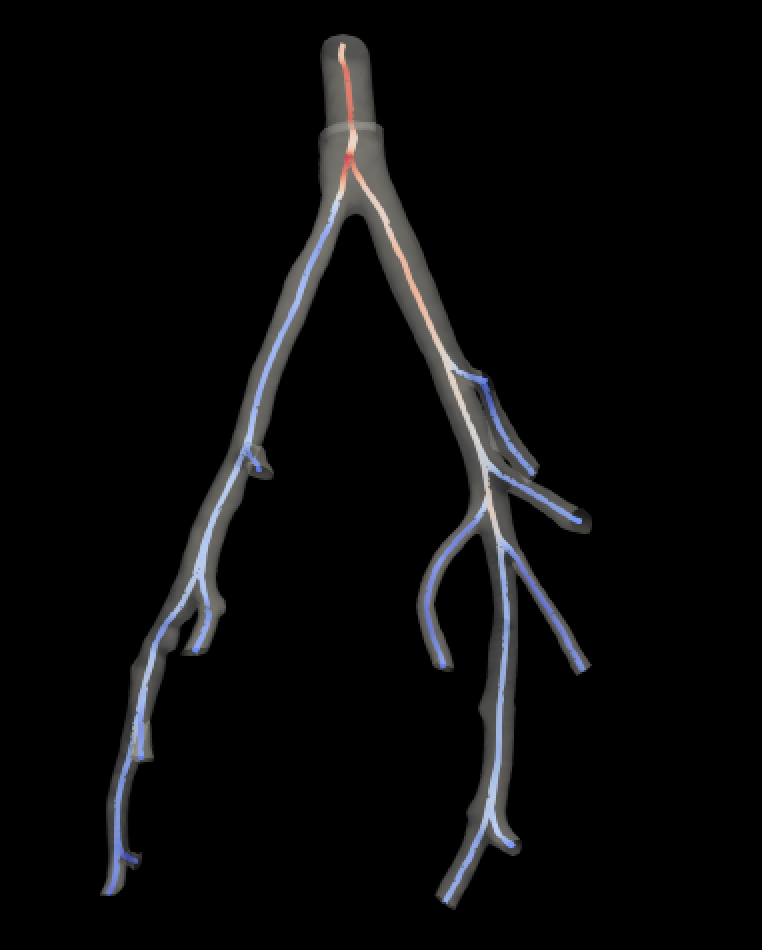}
\includegraphics[scale=0.6]{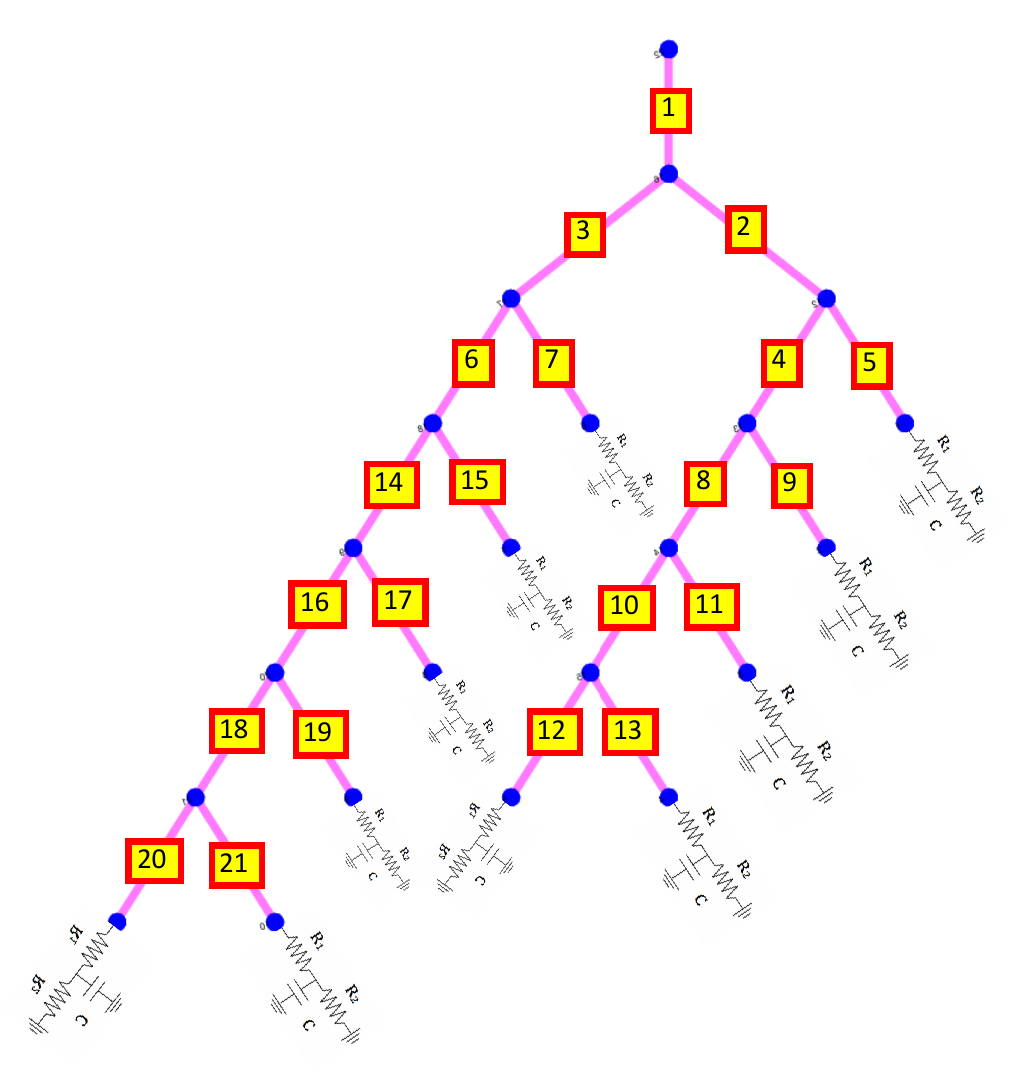}
\caption{3D smoothed network from a micro-CT image of a healthy mouse lung (left) and the connectivity graph of of the same network (right). A similar network was extracted for the hypoxic mouse (not shown here) with the same connectivity graph. Note, the connectivity graph only shows the network structure and does not indicate length and radii of individual vessels. A three-element Windkessel model  with two resistors and a capacitor is attached at the end of each terminal vessel. Detailed network dimensions for healthy and hypoxic mouse can be can be found in \cite{NewPaper}.}
\label{fig:1}
\end{figure}

The flow in the network is driven by prescribing a flow waveform measured in the main pulmonary artery (MPA) of a healthy and hypoxic mouse. Conservation of blood flow ($q_p = q_{d_1} + q_{d_2}$) and continuity of pressure ($p_p = p_{d_1} = p_{d_2}$) are enforced at the bifurcations ($p$ - parent, $d_1$ and $d_2$ - daughters), and an outflow condition predicting vascular bed impedance is prescribed using 3-element Windkessel model, with two resistors $R_1, R_2$ and a capacitor $C$, of the form
\begin{equation}{\label{eq:3}}
   Z(\omega) = R_1 + \frac{R_2}{1+i\omega C R_2} \implies p(L,t) = \frac{1}{T} \int_{0}^{T} q(L, t-\tau)Z(\tau)d\tau, 
\end{equation}
The system of equations (\ref{eq:1}) and (\ref{eq:2}) were  solved numerically using the two step Lax-Wendroff scheme \citep{LaxWendroff}. Simulations were set up to fit pressure waveforms measured in the MPA in a healthy and hypoxic mouse. 

The coupled model contains a total of 100 parameters, used to determine pressure and flow in the network, associated with the vessel fluid dynamics (24 parameters), the geometry (43 parameters), and the outflow boundary conditions (33 parameters). Of these, 45 parameters describing physical properties are fixed. The remaining 55 parameters characterize vessel stiffness (21 parameters - one per vessel), outflow boundary conditions (33 parameters, 3 parameters per terminal vessel), and a tapering factor.  

To reduce the dimension of the parameter space, we assume that the vessel stiffness $f$ is constant throughout the network \citep{Krenz} (reducing the dimension by 20). Second, as described in \cite{NewPaper}, we predict the nominal outflow boundary condition for each vessel from estimates of total resistance and compliance and use a constant scaling factor to optimize each of the resistances and capacitances in the Windkessel models (reducing the dimension by 30). Briefly, total system resistance $R_T \equiv R_1 + R_2 = \overline{p}/\overline{q}$, where $R_1 = 0.2 R_T$ and total capacitance $C_T$ is estimated from diastolic pressure decay. These quantities are distributed to each terminal vessel, indexed $j$, as described in \cite{NewPaper}, giving a nominal set of parameters ${\theta}_{0\text{wk}}^j = \{R^j_{01},R^j_{02},C^j_{0}\}$. Introducing the scaling factors $r_1,r_2$ and $c$, final Windkessel parameters are estimated by optimizing these scaling factors as 
\begin{equation}\label{eq:13}
R^j_1 = (1- 0.5 r_1)R^j_{01}, \quad R^j_2 = (1- 0.5 r_2)R^j_{02}\quad\text{and}\quad C^j = (1- 0.5c)C^j_0,
\end{equation}
To account for the physiologically observed tapering in large arteries, a tapering factor $\xi$ is introduced, such that
\begin{equation}\label{eq:14}
     r_{\textrm{B}} = r_{\textrm{T}} (1-0.5\xi),
\end{equation}
where $r_{\textrm{T}}$ and $r_{\textrm{B}}$ are the inlet and outlet radii of the vessels.

Above assumptions and criteria has allowed us to reduce the dimension of free parameters from 55 to 5. The parameters to be estimated include: vessel stiffness $f$, scaling factors $r_1,r_2,c$ (equation (\ref{eq:13})) and the tapering factor $\xi$.\\

Two network configurations are analysed:
\begin{align}{\label{eq:12}}
\begin{split}
   &\textrm{Network with straight vessels (4D model):}\qquad \boldsymbol{\theta} = \{f,r_1,r_2,c\}\\
  &\textrm{Network with tapered vessels (5D model):}\quad \boldsymbol{\theta} = \{f,r_1,r_2,c,\xi\} \\
\end{split}
\end{align}

\section{Statistical Model}
We start the analysis by specifying the statistical model:
\begin{align}{\label{eq:4}}
\begin{split}
y^q_i = m_q(x_i; \boldsymbol{\theta}) + \epsilon_{1i}, \\
y^p_i = m_p(x_i; \boldsymbol{\theta}) + \epsilon_{2i}, \\
\end{split}
\end{align}
where 
\begin{itemize}
    \item $\mathbf{y}^q = [y^q_1, y^q_2, ..., y^q_n]$ are the noisy measured flows, and $\mathbf{y}^p = [y^p_1, y^p_2, ... y^p_n]$ are the noisy measured pressures,
    \item $m(.)$ describes the system behaviour that comes from numerically solving the fluids model equations ({\ref{eq:1}}) - ({\ref{eq:2}}), i.e. $m_q(.)$ is the output (simulated) flow and $m_p(.)$ is the output (simulated) pressure,
    \item $\boldsymbol{\theta}$ are the parameters to be inferred from the observed flow and pressure. These relate to the Windkessel model parameters $R_1, R_2, C$ defined in (\ref{eq:3}), to the vessel stiffness introduced in (\ref{eq:2}) and to the lung network geometry.  For a more detailed description of the parameters see Section \ref{sssec:mathsmodel},
    \item $x_i\in\mathbf{x}$ denote other input variables (e.g. inflow into the MPA),
    \item $\boldsymbol{\epsilon}$ are the errors, which we assume are i.i.d following a Gaussian distribution and are different for flow and pressure, due to different measurement techniques.
\end{itemize}

Our objective function is chosen to be the residual sum of squares (S), which measures the deviation of our simulated signal from the measured signal and which, under the Gaussian assumption of the errors, is the negative log likelihood of a Normal distribution. 
In this study, S is calculated for pressure in the main pulmonary artery, and we aim to find the parameters that minimise S.
We can write our objective function for pressure as
\begin{align}{\label{eq:5}}
\begin{split}
    S = (\mathbf{y}^p - m_p(\mathbf{x}; \boldsymbol{\theta}))^2 = \sum_{i=1}^{n} (y^p_i - m_p(x_i; \boldsymbol{\theta}))^2, \\
\end{split}
\end{align}
and the log likelihood of the pressure is
\begin{equation}
\log(L) = -n\log \sqrt{2\pi\sigma^2} - \frac{S}{2\sigma^2}.
\end{equation}

For simplicity, in the remainder of the paper, we will no longer use the $p$ index or the $x_i$ term, but it should be noted that all the calculations are done for pressure, and $\mathbf{x}$ is needed in solving the PDEs.

\section{Statistical analysis methods}
In this paper, we apply Markov Chain Monte Carlo (see \cite{MonteCarlo} for a comprehensive review) to the problem of pulmonary circulation to approximately sample the parameters from the posterior distribution. Parameters are inferred using the Delayed Rejection Adaptive Metropolis (DRAM) algorithm \citep{DRAM} and we compare the performance of the DRAM algorithm with that of the  Adaptive Metropolis (AM) algorithm \citep{AM}.
Since MCMC is too slow to converge when started from random points in parameter space, we first obtain point estimates using the maximum likelihood approach, and these estimates are used as initial values in the DRAM or AM algorithm. The concepts of optimization in a maximum likelihood context and uncertainty assessment of these estimates using MCMC are reviewed in subsections \ref{sssec:mle} and subsubsection \ref{sssec:mcmc}. Our work builds on and further improves on existing literature by allowing for a novel parameter scaling technique in the DRAM algorithm with informative priors for the parameters (see subsubsection \ref{sssec:dramscaled}). Following work in existing studies, we employ formal convergence diagnostics for the DRAM algorithm (Multivariate Potential Scale Reduction Factor, Geweke test, Integrated Autocorrelation Time and Effective Sample Size), as described in \ref{sssec:convdiag}. The last step is to use existing scores for selecting the better model for explaining the data (AIC, AICc, BIC, DIC, WAIC), reviewed in subsection \ref{sssec:modelselection}.

\subsection{Maximum likelihood with nonlinear optimization}\label{sssec:mle}
We first apply nonlinear constraint optimization to find the global or local optimum of the objective function, as this exhibits unimodality in low dimensions (see Results section) and additionally, it is nonlinear in the parameters, which are bounded (see the Simulations section for a justification of the parameter ranges). We apply the Sequential Quadratic Programming (SQP) algorithm, which iteratively generates estimates that converge to a solution of (\ref{eq:6}) by constructing and solving quadratic sub-problems that approximate the nonlinear programming problem defined by (\ref{eq:6}). 
\begin{equation}{\label{eq:6}}
\begin{split}
\text{minimize  \hspace{0.5in}} e(\boldsymbol{\theta}),\boldsymbol{\theta} \in \mathbb{R}^d, \text{where } e(\boldsymbol{\theta}) \text{ is the residual-sum-of-squares, } S \text{ (see (\ref{eq:5}))} \\
&\text{\hspace{-4.15in} subject to \hspace{0.45in}} g(\boldsymbol{\theta}) = 0, g:\mathbb{R}^d \rightarrow \mathbb{R}^t \text{ equality constraints defining relations} \\
&\text{\hspace{-2.1in} between parameters,} \\
& \text{\hspace{-3.2in}} h(\boldsymbol{\theta}) \leq 0, h:\mathbb{R}^d \rightarrow \mathbb{R}^u \text{ inequality constraints defining a} \\
& \text{\hspace{-2.15in} restricted parameter ranges.} 
\end{split}
\end{equation}

The gradient-based method used here involves Taylor series expansion and, so, it requires the function to be twice continuously differentiable in the neighbourhood of the point tested and the Hessian matrix of e to be positive definite
This method has been shown to be a very good tool for solving highly nonlinear, fairly low dimensional problems, where we have equality and inequality constraints and where the function displays unimodality. 
In this algorithm, information about the function to be minimised, $e$, and about the equality and inequality functions $g$ and $h$ is stored in a Lagrangian function: $L(\theta, \Lambda, \Gamma) = e(\theta) + \sum_d \Lambda_d g_d(\theta) + \sum_d \Gamma_d h_d(\theta)$, where $\Lambda$ and $\Gamma$ are the Lagrange multipliers. Next, the Lagrangian function is minimised using Newton's method, due to the nonlinearity of the problem, which calls for numerically iterative methods (finding the critical points that make $\nabla L = 0$ is challenging since some of the derivatives may be highly nonlinear). We can optimise the Lagrangian function instead of the objective because the critical points of the Lagrangian are the same as those of the objective function. The problem is then turned into a quadratic subproblem. At iteration k, the direction, v, in which the algorithm searches the minimum is a solution to the equation:
\begin{equation*}
    \min_v e(\theta_k) + \nabla e(\theta_k)^T v + \frac{1}{2}v^T \nabla_{\theta \theta}^2 L(\theta_k, \Lambda_k, \Gamma_k)v,
\end{equation*}
\begin{equation*}
    \hspace{0.7cm} \text{s.t. } g(\theta_k) + \nabla g(\theta_k)^T v \geq 0 \text{ and } h(\theta_k) + \nabla h(\theta_k)^T v = 0.
\end{equation*}
For more details of this algorithm, see \cite{sqpPaper1} and \cite{sqpPaper2}. 


\subsection{Uncertainty quantification with MCMC}
\subsubsection{Overview of MCMC (DRAM)}\label{sssec:mcmc}
The Metropolis--Hastings (MH) algorithm \citep{mh1953} is one of the most widely used MCMC methods. In this algorithm, new samples are generated from the proposal distribution and are either accepted or rejected depending on how likely it is that these samples come from the target distribution, i.e. accept with probability:
\begin{equation}
    \alpha_1(\boldsymbol{\theta}^{k-1}, \boldsymbol{\theta}^*) = \min \left(1, \frac{\pi(\boldsymbol{\theta}^*) q_1(\boldsymbol{\theta}^*,\boldsymbol{\theta}^{k-1})}{\pi(\boldsymbol{\theta}^{k-1}) q_1(\boldsymbol{\theta}^{k-1}, \boldsymbol{\theta}^*)} \right),
\end{equation}
where $\boldsymbol{\theta}_{k-1}$ is the current point, $\boldsymbol{\theta}^*$ the proposed point, $q_1(\boldsymbol{\theta}^*,\boldsymbol{\theta}^{k-1})$ the proposal distribution from $\boldsymbol{\theta}^{k-1}$ to $\boldsymbol{\theta}^*$ and $\pi(.)$ the stationary distribution. If a symmetric proposal distribution $q_1$ (i.e. $q_1(\boldsymbol{\theta}^*,\boldsymbol{\theta}^{k-1}) = MVN(\boldsymbol{\theta}^{k-1}, V_k)$), is used, then the ratio of the proposal distributions will be 1. 
A poor proposal distribution will result in a high rejection rate. To overcome this, upon rejection of $\boldsymbol{\theta}^*$, instead of retaining $\boldsymbol{\theta}^{k-1}$, one can make a second attempt to move \citep{Tierney}; in the $2^{nd}$ proposal, a different distribution is used (i.e. $q_2(\boldsymbol{\theta}^{*2}, \boldsymbol{\theta}^*, \boldsymbol{\theta}^{k-1}) = MVN(\boldsymbol{\theta}^{k-1}, \beta^2 V_k)$ for a symmetric proposal distribution) and this is allowed to depend on the rejected sample, which can be proved to not destroy the Markovian property of the chain \citep{DR}.
Having started in $\boldsymbol{\theta}^{k-1}$ and after rejecting $\boldsymbol{\theta}^*$, the second proposed point, $\boldsymbol{\theta}^{*2}$ is accepted with probability:
\begin{equation}{\label{eq:7}}
    \alpha_2(\boldsymbol{\theta}^{k-1}, \boldsymbol{\theta}^*, \boldsymbol{\theta}^{*2}) = \min \left(1,\frac{\pi(\boldsymbol{\theta}^{*2}) q_1(\boldsymbol{\theta}^{*2}, \boldsymbol{\theta}^*) q_2(\boldsymbol{\theta}^{*2}, \boldsymbol{\theta}^*, \boldsymbol{\theta}^{k-1})[1-\alpha_1(\boldsymbol{\theta}^{*2}, \boldsymbol{\theta}^*)]}{\pi(\boldsymbol{\theta}^{k-1}) q_1(\boldsymbol{\theta}^{k-1}, \boldsymbol{\theta}^*) q_2(\boldsymbol{\theta}^{k-1}, \boldsymbol{\theta}^*, \boldsymbol{\theta}^{*2})[1-\alpha_1(\boldsymbol{\theta}^{k-1}, \boldsymbol{\theta}^*)]} \right),
\end{equation}

\normalsize
\noindent
The acceptance probability in the $2^{nd}$ stage is calculated in such a way that detailed balance is preserved. The ratio of the $q_2$ proposal distributions will become 1 if the proposal distribution is symmetric.  
The Delayed Rejection (DR) algorithm may have a smaller rejection rate, which implies a smaller asymptotic variance of the estimates compared to the MH algorithm \citep{Mira2001a}. 

In cases where the parameters to be learnt are strongly correlated, the parameter dimensionality is large or the problem is highly nonlinear in the parameters, then running the classical MH algorithm can be inefficient, since the chain will move slowly in the parameter space due to the random walk behaviour of the algorithm, thus taking a long time to approach the target density. A solution to this is to adapt the Gaussian proposal covariance depending on the shape (or spatial orientation) and size of the target distribution, i.e. we adapt the proposal covariance matrix, $V_k$ based on the past chain \citep{AM}. It can be shown that this Adaptive Metropolis (AM) algorithm still produces an ergodic Markov chain \citep{AM}.

The proposal distribution in the AM algorithm is a Gaussian distribution centred at the current point $\boldsymbol{\theta}^{k-1}$ and covariance matrix equal to $V_k = V_k(\boldsymbol{\theta}^0, ... \boldsymbol{\theta}^{k-1})$. The proposal covariance matrix, $V_k = R_k^TR_k$ is adapted at given intervals, after possibly some non-adaptation time $t_{\textrm{ad}}$, according to the following formula:
\begin{equation}\label{eq:8}
    V_k = 
     \begin{cases}
      V_0, & \text{if}\ k \leq t_{\textrm{ad}} \\
      s_d cov(\boldsymbol{\theta}^0, ...,\boldsymbol{\theta}^{k-1})+\epsilon I_d, & \text{if}\ k > t_{\textrm{ad}},
    \end{cases}
\end{equation}
where $V_0$ is the initial proposal covariance matrix, $s_d$ is a parameter that depends on the dimension d of the target ($s_d = 2.4^2/d$ has been shown to be optimal for Gaussian targets in terms of mixing) \citep{AM}, $\epsilon > 0$ is a constant that ensures that $V_k$ does not become singular, and $(\boldsymbol{\theta}^0, ... \boldsymbol{\theta}^{k-1})$ are the past chain samples.

DRAM \citep{DRAM} increases the convergence through the use of the rejected values in the DR step (local adaptation), and of the past chain samples to learn about the posterior distribution through the adapted covariance matrix, in the AM step (global adaptation).

\subsubsection{Delayed Rejection Adaptive Metropolis (DRAM) with parameter scaling} \label{sssec:dramscaled}

A pseudocode of the DRAM algorithm can be found in \cite{UncertaintySmith} (pages 175--176). In practice, we implement the algorithm by making use of the Matlab DRAM toolbox \citep{DRAMtoolbox}. However, the toolbox does not account for the case where the parameters differ by orders of magnitude, and so the code had to be modified to allow scaling the parameters. The pseudocode for DRAM with an informative prior, together with the parameter scaling adjustment can be found in Algorithms \ref{alg:algorithm1}-\ref{alg:algorithm4}.
Here we improve the Algorithms 8.8 and 8.10 in \cite{UncertaintySmith}. We introduce a novel extension to DRAM by adding an informative prior and improving the scaling of the parameters. This improvement concerns the way that the acceptance probabilities are calculated. More precisely, the jump densities should be calculated for the transformed parameters (see step \ref{dramScaled1} in Algorithm \ref{alg:algorithm4}). Using the original parameters can cause the ratio of the jump densities to become very large in exceptional cases, making us falsely accept the proposals. For example, if the starting value, which is also the optimum, is close to the boundaries, the algorithm will often jump outside boundaries. This will make some parameters get very high values (e.g. in the second DR try), and so the density in the numerator will become much higher than the density in the denominator in the jump ratio (see equation (\ref{eq:7})). This, however, does not represent a problem for the first stage acceptance probabilities when a symmetric proposal distribution is used, as the proposals will cancel out in the ratio. Hence, the correction only affects DRAM and DR, while no correction is needed for AM.

\begin{algorithm}[htbp]
   \caption{Delayed Rejection Adaptive Metropolis Algorithm (DRAM) with informative prior (no hyperparameters)}
   \label{alg:algorithm1}
   
   \begin{algorithmic}[1]
   
    \State Initialise design parameters: M: number of chain iterations, $\beta$: scaling factor for proposal covariance matrix in DR, $n_{\textrm{dr}}$: number of DR tries, $t_{\textrm{ad}}$: adaptation interval, $b_{\textrm{s}}$: burn-in scale, $b_\textrm{t}$: burn-in time, $\gamma_\textrm{s}^2$: prior for $\sigma^2$ (error variance), $n_\textrm{s}$: prior accuracy for $\gamma_\textrm{s}^2$.
    
    \State Compute $\boldsymbol{\theta}^0 =\argmin_{\boldsymbol{\theta}} \sum_{i=1}^{n} [y_i - m_i(\boldsymbol{\theta})]^2$, where $a_j\leq \theta_j \leq b_j, j = 1 ... d$, $d$: number of parameters, using nonlinear constraint optimization (SQP algorithm) with 20 overdispersed starting values generated from a Sobol sequence. \label{dram2}
    
    \State Initialise $\sigma_0^2 = \frac{S_0}{n-d}$, where $S_0$: optimised residual-of-squares value, $n$: number of observations.
    
    \State Assume that the prior distribution for $\boldsymbol{\theta}$ is a Multivariate Normal distribution (independent $\theta$s),
     $\boldsymbol{\theta} \sim \mathcal{MVN}(\boldsymbol{\mu}, T) $

    \State Set initial $S_{\textrm{prior}}^0 = \sum_{j=1}^d \frac{(\theta^0_j-\mu_j)^2}{t_j^2} $.
    
    \State Either set the initial covariance matrix $V_0$ to be the Hessian matrix from optimization or set $V_0$ via monitoring of acceptance rate. \\
    Set $R_0$ = \textrm{chol}$(V_0)$, where "chol" stands for the Choleski factorization, as an efficient numerical solution of $V_0 = R_0^TR_0$.
    
    \For{k = 1, ... M}
    
    \LState Sample $\mathbf{u}_k \sim \mathcal{MVN}(\mathbf{0},\mathcal{I})$.
    
    \LState Construct candidate from Gaussian proposal: $\boldsymbol{\theta}^* = \boldsymbol{\theta}^{k-1} + R_{k-1} \boldsymbol{u}_k.$ \label{dram9}
    
    \If {$a_j \leq \theta_j \leq b_j$, for all j} \Comment{outside boundaries} 
    
    \LState Jump to DR Algorithm \ref{alg:algorithm3} for iteration $k$
    \Else
    \LState Calculate
    $S_{\boldsymbol{\theta}^*} = \sum_{i=1}^n [y_i - m_i(\boldsymbol{\theta}^*)]^2,
    S_{\textrm{pri}}^* = \sum_{j=1}^d \frac{(\theta^*_j-\mu_j)^2}{t_j^2}$
    \label{dram13}
    \EndIf
    
    \LState Calculate the acceptance probability,
    
    $\alpha_1(\boldsymbol{\theta}^*, \boldsymbol{\theta}^{k-1}) = \min\left(1, e^{-0.5 \left[\frac{S_{\boldsymbol{\theta}^*} - S_{\boldsymbol{\theta}^{k-1}}}{\sigma^2_{k-1}} + S_{\textrm{pri}}^* - S_{\textrm{pri}}^{k-1}\right]}\right).$ \label{dram15}
    
    \LState Sample $v_{\alpha_1} \sim \mathcal{N}(0,1).$
    
    \If {$v_{\alpha_1} < \alpha_1$}
    \LState $\boldsymbol{\theta}^k = \boldsymbol{\theta}^*, S_{\boldsymbol{\theta}^k} = S_{\boldsymbol{\theta}^*}, S_{\textrm{pri}}^k = S_{\textrm{pri}}^*$ \label{dram18}
    \Else
    \LState Enter DR Algorithm \ref{alg:algorithm3} for iteration $k$
    \EndIf
    
    \LState Update 
    $\sigma_k^2 \sim \textrm{Inv-Gamma} \left(\frac{n_\textrm{s}+n}{2}, \frac{n_\textrm{s} \gamma_\textrm{s}^2+S_{\boldsymbol{\theta}^k}}{2} \right)$
    
    \LState Enter AM Algorithm \ref{alg:algorithm2} for iteration $k$
    
    \EndFor
\end{algorithmic}
\end{algorithm}

\begin{algorithm}[htbp]
   \caption{Adaptive Metropolis Algorithm (AM)}
   \label{alg:algorithm2}
   
   \begin{algorithmic}[1]

    \If {$k < b_\textrm{t}$} \Comment{during burn-in time, no adaptation, just scaling R}
        \LState
        $R_k \leftarrow R_{k-1}/b_\textrm{s}$, \Comment{decrease R if rejection rate $>$ 0.95}
        \LState
        $R_k \leftarrow b_\textrm{s} R_{k-1}$, \Comment{increase R if rejection rate $<$ 0.05}
    \Else
        \If {$mod(k, t_{\textrm{ad}}) = 1$}
            \LState Set 
            $V_k = s_d cov(\boldsymbol{\theta}^0, \boldsymbol{\theta}^1, ... \boldsymbol{\theta}^k) + \epsilon \mathcal{I}_d,
            R_k = chol(V_k)$
        \Else
        \LState
        $V_k = V_{k-1},
        R_k =R_{k-1}$
        \EndIf
    \EndIf
    \State Output $R_k$

\end{algorithmic}
\end{algorithm}

\begin{algorithm}[htbp]
   \caption{Delayed Rejection Algorithm (DR) for $n_{\textrm{dr}}=2$, i.e. for $2^{nd}$ stage proposals}
   \label{alg:algorithm3}
   
   \begin{algorithmic}[1]

    \State Sample $\mathbf{u}_k \sim \mathcal{MVN}(\mathbf{0},\mathcal{I})$.
    
    \State Construct $2^{nd}$ stage candidate from Gaussian proposal: $\boldsymbol{\theta}^{*2} = \boldsymbol{\theta}^{k-1} + \beta R_{k-1} \boldsymbol{u}_k.$ \label{dr2}
    
    \If {$a_j \leq \theta_j \leq b_j$, for all j}
    
    \LState 
    $\boldsymbol{\theta}^k = \boldsymbol{\theta}^{k-1}, S_{\boldsymbol{\theta}^k} = S_{\boldsymbol{\theta}^{k-1}}, 
    S_{\textrm{pri}}^k = S_{\textrm{pri}}^{k-1}$
    
    \LState
    and jump to AM Algorithm \ref{alg:algorithm2}.
    \Else
    
    \LState
    $S_{\boldsymbol{\theta}^{*2}} = \sum_{i=1}^n [y_i - m_i(\boldsymbol{\theta}^{*2})]^2,
    S_{\textrm{pri}}^{*2} = \sum_{j=1}^d \frac{(\theta^{*2}_j-\mu_j)^2}{t_j^2}$
    \label{dr4}
    
    \EndIf
    
    \State Calculate the acceptance probability,
    $ \alpha_2(\boldsymbol{\theta}^{*2}, \boldsymbol{\theta}^*, \boldsymbol{\theta}^{k-1}) $ according to equation (\ref{eq:7}). \label{dr6}
    
    \State Sample $v_{\alpha_2} \sim \mathcal{N}(0,1).$
    
    \If {$v_{\alpha_2} < \alpha_2$},
    
    \LState
    $\boldsymbol{\theta}^k = \boldsymbol{\theta}^{*2}, S_{\boldsymbol{\theta}^k} = S_{\boldsymbol{\theta}^{*2}}, S_{\textrm{pri}}^k = S_{\textrm{pri}}^{*2}$ \label{dr8}
    \Else
    \LState
    $\boldsymbol{\theta}^k = \boldsymbol{\theta}^{k-1}, S_{\boldsymbol{\theta}^k} = S_{\boldsymbol{\theta}^{k-1}}, S_{\textrm{pri}}^k = S_{\textrm{pri}}^{k-1}$ \label{dr9}
    \EndIf

\end{algorithmic}
\end{algorithm}

\begin{algorithm}[ht]
   \caption{Delayed Rejection Adaptive Metropolis (DRAM) with scaled parameters}
   \label{alg:algorithm4}
   
   \begin{algorithmic}[1]
   \State Denote $\boldsymbol{\theta}_s = \boldsymbol{\theta}./\mathbf{s}$, where $\mathbf{s}$ are the scaling factors.

    \State In Algorithm \ref{alg:algorithm1}:
    \State DRAM \ref{dram2}: Compute $\boldsymbol{\theta}^0 =\argmin_{\boldsymbol{\theta}} \sum_{i=1}^{n} [y_i - m_i(\boldsymbol{\theta}.\times \mathbf{s})]^2$

    \State DRAM \ref{dram9}: Construct candidate from Gaussian proposal:
    $$\boldsymbol{\theta}_s^* = \boldsymbol{\theta}_s^{k-1} + R_{k-1} \boldsymbol{u}_k.$$

    and set $$\boldsymbol{\theta}^* = \boldsymbol{\theta}_s^* .\times \mathbf{s}$$

    \State DRAM \ref{dram13}: Set
    $$S_{\textrm{pri}}^* = \sum_{j=1}^d \frac{(\theta^*_j-\mu_j)^2}{t_j^2} + \log(det(J))$$
    where $J$ is the Jacobian of the parameter transformation. If the parameters are assumed independent and $\mathbf{s}$ is constant, then $\log(det(J))=\sum_{j=1}^d s_j$.  

    \State DRAM \ref{dram15}: Calculate the acceptance probability,
    $$ \alpha_1(\boldsymbol{\theta}_s^*, \boldsymbol{\theta}_s^{k-1})$$
    
    \State \textbf{Note}: if a noninformative prior is used and $\boldsymbol{s}$ is a constant, then we are left with the likelihood contribution in the acceptance probability, i.e. $\alpha = \min\left(1, e^{-0.5 \left[\frac{S_{\boldsymbol{\theta}^*} - S_{\boldsymbol{\theta}^{k-1}}}{\sigma^2_{k-1}}\right]}\right)$.
    In addition, in the likelihood calculations, we always use the unscaled parameters, whereas the priors and proposals are done in the transformed parameter space.

    \State DRAM \ref{dram18}: In addition, set $\boldsymbol{\theta}_s^k = \boldsymbol{\theta}_s^*.$

    \State In Algorithm \ref{alg:algorithm3}:

    \State DR \ref{dr2}: Construct $2^{nd}$ stage candidate from Gaussian proposal: $$\boldsymbol{\theta}_s^{*2} = \boldsymbol{\theta}_s^{k-1} + \beta R_{k-1} \boldsymbol{u}_k,$$
    and set
    $$\boldsymbol{\theta}^{*2} = \boldsymbol{\theta}_s^{*2} .\times \mathbf{s}$$

    \State DR \ref{dr4}: Set
    $$S_{\textrm{pri}}^{*2} = \sum_{j=1}^d \frac{(\theta^{*2}_j-\mu_j)^2}{t_j^2} + \log(det(J))$$

    \State DR \ref{dr6}: Calculate the acceptance probability in the scaled parameter space,
    $ \alpha_2(\boldsymbol{\theta}_s^{*2}, \boldsymbol{\theta}_s^*, \boldsymbol{\theta}_s^{k-1}) $. \label{dramScaled1}

    \State DR \ref{dr8} or DR \ref{dr9}: In addition, set 
    $\boldsymbol{\theta}_s^k = \boldsymbol{\theta}_s^{*2}$ 
    or 
    $\boldsymbol{\theta}_s^k = \boldsymbol{\theta}_s^{k-1}$.

\end{algorithmic}
\end{algorithm}

\clearpage
\textbf{Note}: If a noninformative prior is used, then $S_{\textrm{pri}} = 0$ in Algorithms \ref{alg:algorithm1}, \ref{alg:algorithm3} and \ref{alg:algorithm4}.

\subsection{Convergence Diagnostics}\label{sssec:convdiag}
Once the Markov chains are produced using the algorithms discussed above, we test for convergence, i.e. test whether our Markov chains have converged in distribution to the posterior distribution of interest.

\subsubsection{Integrated Autocorrelation Time (IACT)}
One way to assess convergence is to assess the autocorrelations between the draws of the Markov chain. The lag $l$ autocorrelation $\rho_l$ is the correlation between every draw and its $l^{\textrm{th}}$ lag:
\begin{equation}
\rho_l = \frac{\sum_{k=1}^{M-l}(\theta_k - \bar{\theta})(\theta_{k+l}-\bar{\theta})}{\sum_{k=1}^{M}(\theta_k-\bar{\theta})^2}
\end{equation}

The Integrated Autocorrelation Time \citep{IACT} is given by:
$\tau = 1 + 2\sum_{l=1}^{\infty}\rho_l$.

It would be expected that the $l^{\textrm{th}}$ lag autocorrelation is smaller as $l$ increases ($2^{\textrm{nd}}$ and $50^{\textrm{th}}$ draws should be less correlated than $2^{\textrm{nd}}$ and $4^{\textrm{th}}$ draws). Relatively high autocorrelation for higher values of $l$ suggests high degree of correlation between draws and thus, slow mixing.

Having IACT, we can also derive the Effective Sample Size (ESS) \citep{ESS}, which gives an estimate of the equivalent number of independent iterations that the chain represents. Low ESS can indicate that the sampler used is inefficient. 

The formula for ESS is given by:
\begin{equation}
ESS = \frac{M}{1 + 2\sum_{l=1}^{\infty}\rho_l} = \frac{M}{\tau}.
\end{equation}


\subsubsection{Geweke test}

The Geweke test \citep{Geweke} formally tests for equality of the mean of two sub-chains (typically the first $10\%$ and last $50\%$ of the iterations). A low p-value for the Z-statistic will indicate non-convergence.

\subsubsection{Brooks Gelman Rubin test}

The Brooks Multivariate Potential Scale Reduction Factor (MPSRF) \citep{BrooksGelman} can be calculated for multiple chains from starting values overdispersed with respect to the stationary distribution. MPSRF is a multivariate extension to the Gelman-Rubin test \citep{GelmanRubin}, assessing convergence of the parameters simultaneously. Not only does this take into account the variances of individual parameters, but also the interactions between the parameters (the covariances). The MPSRF is calculated in such a way that it is an approximate estimate of the maximum upper bound of the univariate SRF for every parameter.
Suppose we have $m$ chains in parallel, and $\boldsymbol{\theta}^k_i$ is the set of parameters drawn at the $k^{\textrm{th}}$ iteration, $k = 1, ...M$ from the $i^{\textrm{th}}, i = 1, ...m$ chain. Then we can calculate the posterior variance-covariance matrix, as follows:
\begin{equation*}
    \hat{V} = \frac{M-1}{M} W + \left( 1+\frac{1}{m}\right)\frac{B}{M},
\end{equation*}
where
\begin{equation*}
    W = \frac{1}{m(M-1)}\sum_{i=1}^m \sum_{k=1}^M (\boldsymbol{\theta}_i^k - \bar{\boldsymbol{\theta}}_{i.}) (\boldsymbol{\theta}_i^k - \bar{\boldsymbol{\theta}}_{i.})^{'}
\end{equation*}
and
\begin{equation*}
    B = \frac{M}{m-1}\sum_{i=1}^m (\bar{\boldsymbol{\theta}}_{i.} - \bar{\boldsymbol{\theta}}_{..}) (\bar{\boldsymbol{\theta}}_{i.} - \bar{\boldsymbol{\theta}}_{..})^{'}
\end{equation*}
$W$ is the within-chain covariance matrix and $B$ the between-chain covariance matrix. If $\lambda_1$ is the largest eigenvalue of the symmetric and positive definite matrix $W^{-1}B$ (see proof in \citep{BrooksGelman}), then the MPSRF,
\begin{equation*}
    \hat{R}^p = \frac{M-1}{M} + \frac{m+1}{m} \lambda_1
\end{equation*}
approaches 1 if $M \rightarrow \infty$ and the chains have converged to the stationary distribution, i.e. $W \approx B$ and, hence $\lambda_1 \rightarrow 0$.

\noindent
In addition, a MPSRF plot is useful in indicating whether the lack of convergence is due to the correlation between parameters.

\subsection{Model selection}\label{sssec:modelselection}
Like in any other Bayesian application, where we have two or more competing models, we may be interested in comparing them, in order to choose the better model. A more complex model is not necessarily favoured, despite the fact that it provides a better fit. Penalising for complexity (too many parameters) is a necessary adjustment to prevent from choosing an overparameterised model that overfits the data and is not generalisable to future data. In this paper, we use the  corrected Akaike Information Criterion (AICc) \citep{AICc}, the Bayesian Information Criteria (BIC) \citep{BIC}, the Deviance Information Criteria (DIC) \citep{DIC} and the Watanabe AIC (WAIC) \citep{WAIC2010} score for model selection, and the model with the lowest score will be preferred.

\subsubsection{AICc and BIC}

AICc \citep{AICc} and BIC \citep{BIC} can be used for model selection in the maximum likelihood framework, where we obtain a maximum likelihood estimate. We use the maximum log likelihood (i.e. the minimum residual sum-of-squares, S) to represent the goodness of fit, and we apply a penalty to avoid overfitting. AICc and BIC penalise for the model complexity differently: AICc applies a lower penalty, while BIC tends to penalise more and choose simpler models.
AICc is a modification of AIC \citep{AIC} by adding a correction for finite sample sizes, hence applying a higher penalty to avoid overfitting. The formula for AICc is given for a univariate Gaussian errors model:
\begin{align}
\begin{split}
  AICc &= -2\log(\mathbf{y}|\hat{\boldsymbol{\theta}}_{\textrm{mle}}) + 2d + \frac{2d(d+1)}{n-d-1}, \\
  BIC &= -2\log(\mathbf{y}|\hat{\boldsymbol{\theta}}_{\textrm{mle}}) + d\log n,
\end{split}
\end{align}
where $\log(\mathbf{y}|\hat{\boldsymbol{\theta}}_{\textrm{mle}})$ is the maximum log-likelihood, $d$ is the number of parameters in the model and $n$ is the total number of observations.
We can derive these formulas to depend on the residual sum-of-squares and the formula involving $S$ is given in equation (\ref{eq:9}):
\begin{align}
\label{eq:9}
\begin{split}
  AICc &= \frac{S}{\sigma^2} + n\log \sigma^2 + 2d + \frac{2d(d+1)}{n-d-1} + C,\\
  BIC &= \frac{S}{\sigma^2} + n\log \sigma^2 + d\log n + C,
\end{split}
\end{align}
where $C = n\log 2\pi$ is a constant.

\subsubsection{DIC}
DIC \citep{DIC} is a partially Bayesian version of AIC, where $\hat{\boldsymbol{\theta}}_{\textrm{mle}}$ is replaced by the posterior mean $\hat{\boldsymbol{\theta}}_{\textrm{Bayes}} = E(\boldsymbol{\theta}|\mathbf{y})$ and the penalisation is data driven. 
Then,
\begin{equation}
    DIC = -2\log p(\mathbf{y}|\hat{\boldsymbol{\theta}}_{\textrm{Bayes}}) + 2p_{\textrm{DIC}},
\end{equation}
where
\begin{equation} \label {eq:10}
    p_{\textrm{DIC}} = 2 \left(\log p(\mathbf{y}|\hat{\boldsymbol{\theta}}_{\textrm{Bayes}}) - E_{\boldsymbol{\theta}|\mathbf{y}}(\log p(\mathbf{y}|\boldsymbol{\theta}))\right),
\end{equation}
where $E_{\boldsymbol{\theta}|\mathbf{y}}(\log(\mathbf{y}|\boldsymbol{\theta}))$ is the expectation of the log posterior predictive density,
which in practice is calculated by replacing the expectation with the average over the posterior draws, as follows \citep{gelman2013}:
\begin{equation}
    \text{computed } p_{\textrm{DIC}} = 2 \left(\log p(\mathbf{y}|\hat{\boldsymbol{\theta}}_{\textrm{Bayes}}) - \frac{1}{S}\sum_{s=1}^S \log p(\mathbf{y}|\boldsymbol{\theta}^s)\right),
\end{equation}
where the sum is over the set of parameters $\boldsymbol{\theta}^s, s = 1, ...S$, that have approximately been drawn from the posterior distribution with MCMC (i.e. $\boldsymbol{\theta}^s$ is the MCMC sample).

\subsubsection{WAIC}

WAIC \citep{WAIC2010} is fully Bayesian in the sense that it is computed using the whole posterior distribution. It takes the computed log posterior predictive density in equation (\ref{eq:11}) and adjusts for overfitting by adding a correction for the effective number of parameters \citep{gelman2013}.
\begin{equation}
    WAIC = -2 lppd + 2 p_{\textrm{WAIC}},
\end{equation}
where
\begin{equation}\label{eq:50}
    lppd = log \prod_{i=1}^n p_{\boldsymbol{\theta}|\mathbf{y}}(y_i) =  \sum_{i=1}^n \log \int p(y_i|\boldsymbol{\theta})p(\boldsymbol{\theta}|\mathbf{y}) d\boldsymbol{\theta}
\end{equation}
The left hand side term of equation (\ref{eq:50}), $p_{\boldsymbol{\theta}|\mathbf{y}}(y_i)$ equals the integral on the right hand side because $\boldsymbol{\theta}$ is unknown, and so we take an expectation of $p_{\boldsymbol{\theta}|\mathbf{y}}(y_i)$ over $\boldsymbol{\theta}$ given $\mathbf{y}$, i.e. we integrate out $\boldsymbol{\theta}$.
In practice, this is computed as follows:
\begin{equation} \label{eq:11}
    \text{computed } lppd = \sum_{i=1}^n \log \left (\frac{1}{S} \sum_{s=1}^S p(y_i|\boldsymbol{\theta}^s)\right)
\end{equation}
where $\boldsymbol{\theta}^s, s = 1, ...S$, are posterior samples from $p(\boldsymbol{\theta}|\mathbf{y})$,
and
\begin{equation}
    p_{\textrm{WAIC}} = \sum_{i=1}^n var_{\boldsymbol{\theta}|\mathbf{y}} (\log p(y_i|\boldsymbol{\theta}))
\end{equation}
In practice, to compute the posterior variance  for every term in the log predictive density $V_{s=1}^S \log p(y_i|\boldsymbol{\theta}^s)$, where $V_{s=1}^S$ is the sample variance, we use: $V_{s=1}^S a_s = \frac{1}{S-1}\sum_{s=1}^S (a_s - \bar{a})^2.$
The effective number of parameters is then given by the sum of $V_{s=1}^S$ over all the data points:
\begin{equation}
    \text{computed } p_{\textrm{WAIC}} = \sum_{i=1}^n V_{s=1}^S (\log p(y_i|\boldsymbol{\theta}^s)).
\end{equation}

AIC, AICc, DIC, WAIC can be used for model selection through predictive accuracy, whereas BIC is used to approximate the marginal probability density of the data under the model $p(\mathbf{y})$, useful for discrete model selection when estimating the relative posterior probabilities \citep{gelman2013}.

Out of AIC, AICc, DIC, WAIC, the latter one is preferred as it has better asymptotic properties. This is because in their derivation using the Taylor series expansion of the information distance between the model and the real system around the optimal parameter that minimises this distance, terms above the third order are neglected for WAIC, whereas for AIC for example, terms above second order are neglected, so the error associated with WAIC is lower, i.e. the predictive accuracy is higher for WAIC \citep{WAIC2015}. In addition, WAIC is asymptotically equivalent to Bayesian leave-one-out cross-validation (LOO-CV) \citep{WAIC2010}. In Bayesian Loo- CV, one point is removed at a time from the data, the model is trained on the $n-1$ observations and the model fit to the data assessed by computing the log predictive density of the holdout data. A model that has a higher Bayesian LOO- CV estimate of the out-of-sample predictive fit (or lower -2 times this estimate, to be on the deviance scale) is preferred. 
The Baysian LOO-CV estimate of the out-of-sample predictive fit is:
\begin{equation*}
    lppd_{LOO-CV} = \sum_{i=1}^n \log p(y_i|\mathbf{y}_{-i}) = \sum_{i=1}^n \log \int p(y_i|\boldsymbol{\theta})p(\boldsymbol{\theta}|\mathbf{y}_{-i})d\boldsymbol{\theta},
\end{equation*}
which in practice is calculated as:
\begin{equation*}
    \text{computed } lppd_{LOO-CV} = \sum_{i=1}^n \log \left(\frac{1}{S}\sum_{s=1}^S p(y_i|\boldsymbol{\theta}^s)\right),
\end{equation*}
where ${\boldsymbol{\theta}^s}$ is a sample from $p(\boldsymbol{\theta}|\mathbf{y}_{-i})$.

AIC, AICc, DIC and WAIC are scores for the adjusted within-sample predictive accuracy, i.e. all points in $\mathbf{y}$ are used to calculate the scores and prediction is done for the same points; the bias associated with doing this goes to zero as the number of observations increases, i.e. asymptotically. The scores use a penalty term for the model complexity. In contrast, LOO-CV measures the out-of-sample prediction accuracy, as its calculation relies on the points in $\mathbf{y}_{-i}$, i.e. on all points, except the $i^{\textrm{th}}$, and we calculate the predictive accuracy for the $i^{\textrm{th}}$ point.
What is more, when comparing WAIC and DIC, WAIC can produce a reliable result even when dealing with multimodal posterior distributions, as it uses the entire posterior distribution, while DIC would be misleading as it uses the mean posterior estimate (e.g. in a bimodal posterior distribution, the mean would be between the two modes). It is known that DIC cannot deal with singular models \citep{WAIC2010}. 

\section{Simulations}
The parameters that we wish to infer, described in detail in Section 2 are taken as arguments to the PDEs (\ref{eq:1}) - (\ref{eq:2}), and predicted flow and pressure are obtained along several locations in the 21 vessels. Since measured data are only available from one location in the main pulmonary artery, we only use the corresponding prediction from the MPA. We drive the system by the measured flow data, and compare our predictions to pressure measurements in the MPA. This is done using the residual sum-of-squares, S (equation (\ref{eq:5})). The set of parameters that give the closest prediction to the measured data is our estimate (maximum likelihood estimate) \citep{PaunIWSM}. In practice, this is performed using the nonlinear constrained optimization scheme (SQP algorithm). The uncertainty around these estimates is obtained by approximately sampling from the posterior distribution using MCMC (DRAM, AM). We implement these algorithms by making use of the DRAM toolbox implemented in Matlab \citep{DRAMtoolbox}.
 
The parameters are on different scales: $f \in [10^4, 10^6], r_1, r_2, c \in [-3, 2.5], \xi \in [0, 0.5]$, and these parameter bounds are chosen to be biologically meaningful. To avoid having an ill-conditioned problem induced by a high condition number in the Hessian matrix \citep{Illcond}, we rescale the parameters to have the same order of magnitude and use these scaled parameters in the optimization routine, as well as in the DRAM algorithm. 
There are certain parameter configurations outside the domain of convergence of the numerical scheme. This prompted us to assign a very high value to the RSS $(10^{10})$ that marks the unsuccessful simulations (data are unlikely given those parameter values).

The 20 different initial values for the parameters passed in the optimization algorithm are uniformly drawn from a Sobol sequence to ensure a good coverage of the multidimensional parameter space \citep{Sobol}. The SQP algorithm is iterated until it satisfies the convergence criterion, i.e. $|\boldsymbol{\theta_{i}} - \boldsymbol{\theta_{i+1}}| < 1e-11$. Once convergence has been reached, we take the optimum parameter values as starting values for the DRAM algorithm in order to speed up the simulations.

Let us now define the likelihood, priors, proposal distribution and posterior distribution used in the DRAM algorithm.
\begin{itemize}
    \item Data likelihood: $y_i|\boldsymbol{\theta} \sim N(m_i(\boldsymbol{\theta}), \sigma^2)$, i.e.
    \begin{equation*}
        p(\mathbf{y}|\boldsymbol{\theta}, \sigma^2) = \left(\frac{1}{\sqrt{2\pi\sigma^2}}\right)^n \exp \left(-\frac{\sum_{i=1}^n(y_i-m_i(\boldsymbol{\theta}))^2}{2\sigma^2} \right)
    \end{equation*}
    \item Priors: $\boldsymbol{\theta} \sim \text{Truncated } MVN(\boldsymbol{\mu}, T), T: \text{positive definite matrix}$, and due to lack of prior knowledge, we assign high variances to the parameters, to make the prior noninformative.
    \\
    $\sigma^2 \sim \textrm{Inv-Gamma}(a,b)$, where following Algorithm \ref{alg:algorithm1}, we denote $a = \frac{n_s}{2}>0, b = \frac{n_s\gamma_s^2}{2}>0$ to be the shape and scale parameters, with $\gamma_s^2$ being the prior for $\sigma^2$ and $n_s$ the prior accuracy for $\gamma_s^2$ -- a large value of $n_s$ means we are certain of our prior, i.e. the prior dominates, and so the posterior samples will be close to $\sigma_s^2$. The reason for choosing this prior is that the Inv-Gamma family is conditionally conjugate (i.e. if the prior distribution for $\sigma^2$ is an Inv-Gamma, then the conditional posterior distribution for $\sigma^2$ is also Inv-Gamma), allowing us to easily draw posterior samples for $\sigma^2$ from its full conditional distribution in a Gibbs step. We assume a noninformative prior for $\sigma^2$, so we set $n_s = 1$ and let $\gamma_s^2$ be the initial noise variance, $\sigma_2$ chosen by the user (the defaults in the toolbox).  
    \begin{align*}
    \begin{split}
        p(\boldsymbol{\theta}) = \det(2\pi T)^{-\frac{1}{2}} \exp \left(-\frac{1}{2}(\boldsymbol{\theta}-\boldsymbol{\mu})^{'}T^{-1}(\boldsymbol{\theta}-\boldsymbol{\mu})\right), \\
        p(\sigma^2) = \frac{b^a}{\Gamma(a)} (\sigma^2)^{-a-1} \exp \left(-\frac{b}{\sigma^2}\right)
    \end{split}
    \end{align*}
    \item Proposal distribution: $q(\boldsymbol{\theta}^*, \boldsymbol{\theta}^{k-1}) = \text{Truncated } MVN(\boldsymbol{\theta}^{k-1}, V)$.
    
    For $\boldsymbol{\theta}^*$ inside boundaries:
    \begin{equation*}
        q(\boldsymbol{\theta}^*,\boldsymbol{\theta}^{k-1}) = \det(2\pi V)^{-\frac{1}{2}} \exp \left(-\frac{1}{2}(\boldsymbol{\theta}^*-\boldsymbol{\theta}^{k-1})^{'}V^{-1}(\boldsymbol{\theta}^*-\boldsymbol{\theta}^{k-1})\right), \\
    \end{equation*}
 
    \item $\text{Posterior distribution} \propto \text{Likelihood} \times \text{Prior}$, i.e.
    \begin{equation*}
        p(\boldsymbol{\theta}, \sigma^2|\mathbf{y}) \propto p(\mathbf{y}|\boldsymbol{\theta}, \sigma^2) p(\boldsymbol{\theta}) p(\sigma^2)
    \end{equation*}
    which yields:
    \begin{align*}
    \begin{split}
        p(\boldsymbol{\theta}, \sigma^2|\mathbf{y}) \propto (\sigma^2)^{-\frac{n}{2}-a-1} \exp \left(-\frac{0.5\sum_{i=1}^n (y_i-m_i(\boldsymbol{\theta}))^2 + b}{\sigma^2}\right) \times \\
        \exp \left(-0.5 (\boldsymbol{\theta}-\boldsymbol{\mu})^{'}T^{-1}(\boldsymbol{\theta}-\boldsymbol{\mu})\right).
    \end{split}
    \end{align*}
    
    In this expression, $\mathbf{m}(\boldsymbol{\theta})$ is the prediction from the PDEs and $\mathbf{y}$ is the measured data.
\end{itemize}

DRAM is used to sample the parameters $\boldsymbol{\theta}$ in a Metropolis-Hastings step. We then fix $\boldsymbol{\theta}$ to the sample drawn, and we update $\sigma^2$, the error variance, in a Gibbs step, by drawing samples from an Inverse-Gamma distribution, $\textrm{Inv-Gamma}\left(\frac{n}{2}+a, 0.5\sum_{i=1}^n (y_i-m_i(\boldsymbol{\theta}))^2 + b\right) = \textrm{Inv-Gamma}\left(\frac{n_s+n}{2}, \frac{n_s\gamma_s^2 + \sum_{i=1}^n (y_i-m_i(\boldsymbol{\theta}))^2}{2}\right)$. We notice that when going from the prior to the posterior, the distribution is modified to include the contribution from the data, $\frac{\sum_{i=1}^n (y_i-m_i(\boldsymbol{\theta}))^2}{2}$, and the accuracy in the $\sigma^2$ sample increases when data become available (we add $\frac{n}{2}$ to the prior shape parameter - the more data points, the higher the accuracy).
We use the $\boldsymbol{\theta}$ samples as an input in the mathematical equations (\ref{eq:1}) - (\ref{eq:2}), and thus, for every DRAM sample we solve the PDEs. Since every forward simulation takes around 20s of CPU time, in the interest of time, DRAM is started from optimised values. In our analysis, we compare the pressure prediction obtained using the DRAM samples to the measured pressure.

To check for convergence of the Markov chains, we have employed the Geweke test, the Integrated Autocorrelation Time, the Effective Sample Size, and the Brooks Gelman Rubin test for sub-chains of the same chain started from optimised values. We acknowledge the fact that we do not satisfy a fundamental condition of this test (using parallel chains started from overdispersed values), but to make the method computationally viable, adopting these heuristics is indispensable.

Regarding model selection, we have compared the model with tapering (5D) to the model without the tapering factor (4D). 

\section{Results}
This analysis is done on two mice, a control and a hypoxic mouse. For each mouse, we aim to compare two models: a 5D model which includes all the parameters considered (see (\ref{eq:12})), and a 4D model (nested within the 5D model) which excludes the tapering factor $\xi$, as it is uncertain whether it is needed in the model.

\subsection{Optimization Results}
\begin{figure}[h!]
\centering
\subfigure[Residual sum-of-squares (S) for parameters $f$ and $r_2$ with $S: (1.2\times10^2, 6.5\times10^4)$]{\label{fig:2a}\includegraphics[width=0.75\textwidth]{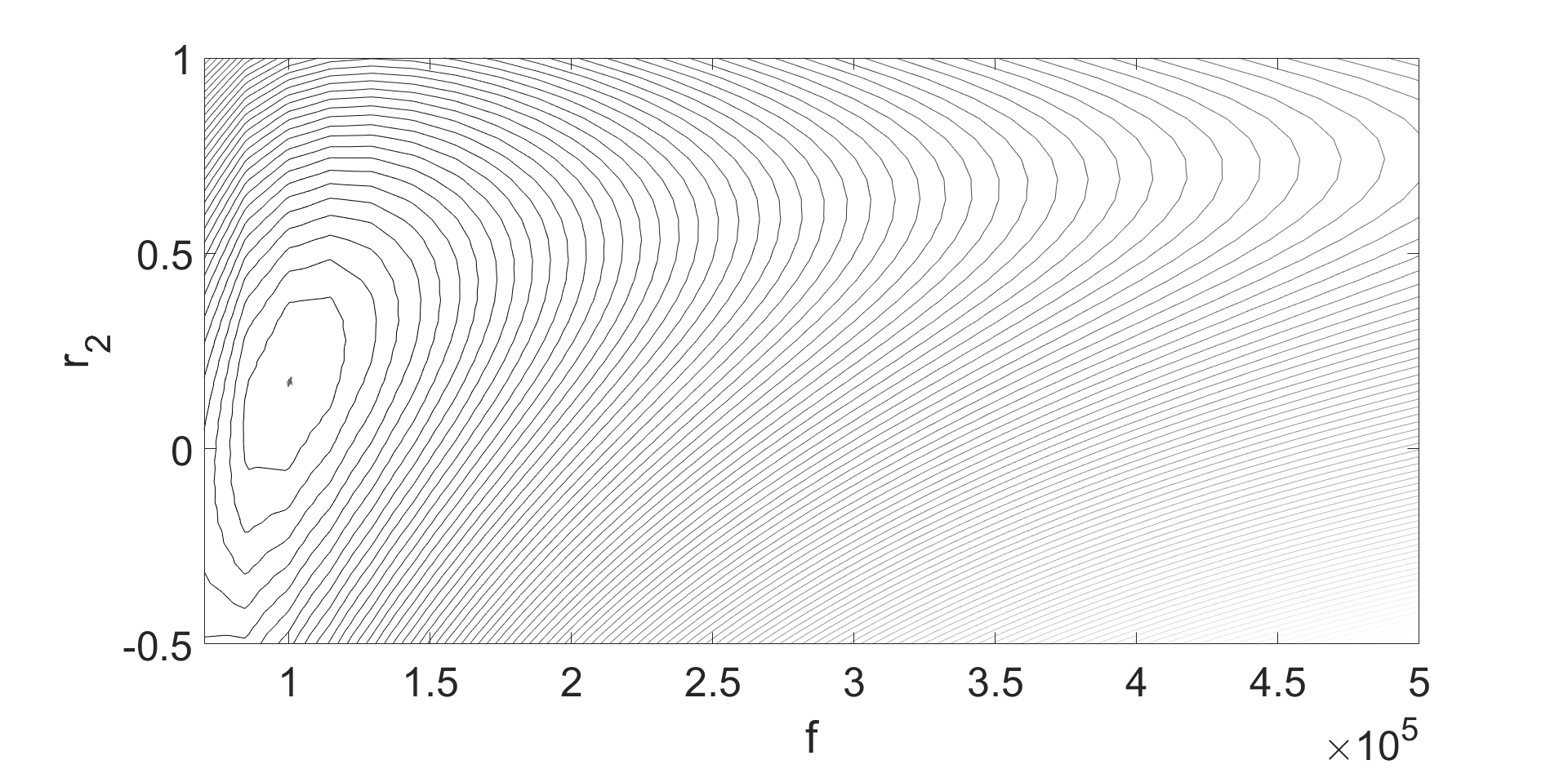}}
\subfigure[Residual sum-of-squares (S) for parameters $r_2$ and $c$ with $S: (1.2\times10^2, 2.2\times10^4)$]{\label{fig:2c}\includegraphics[width=0.75\textwidth]{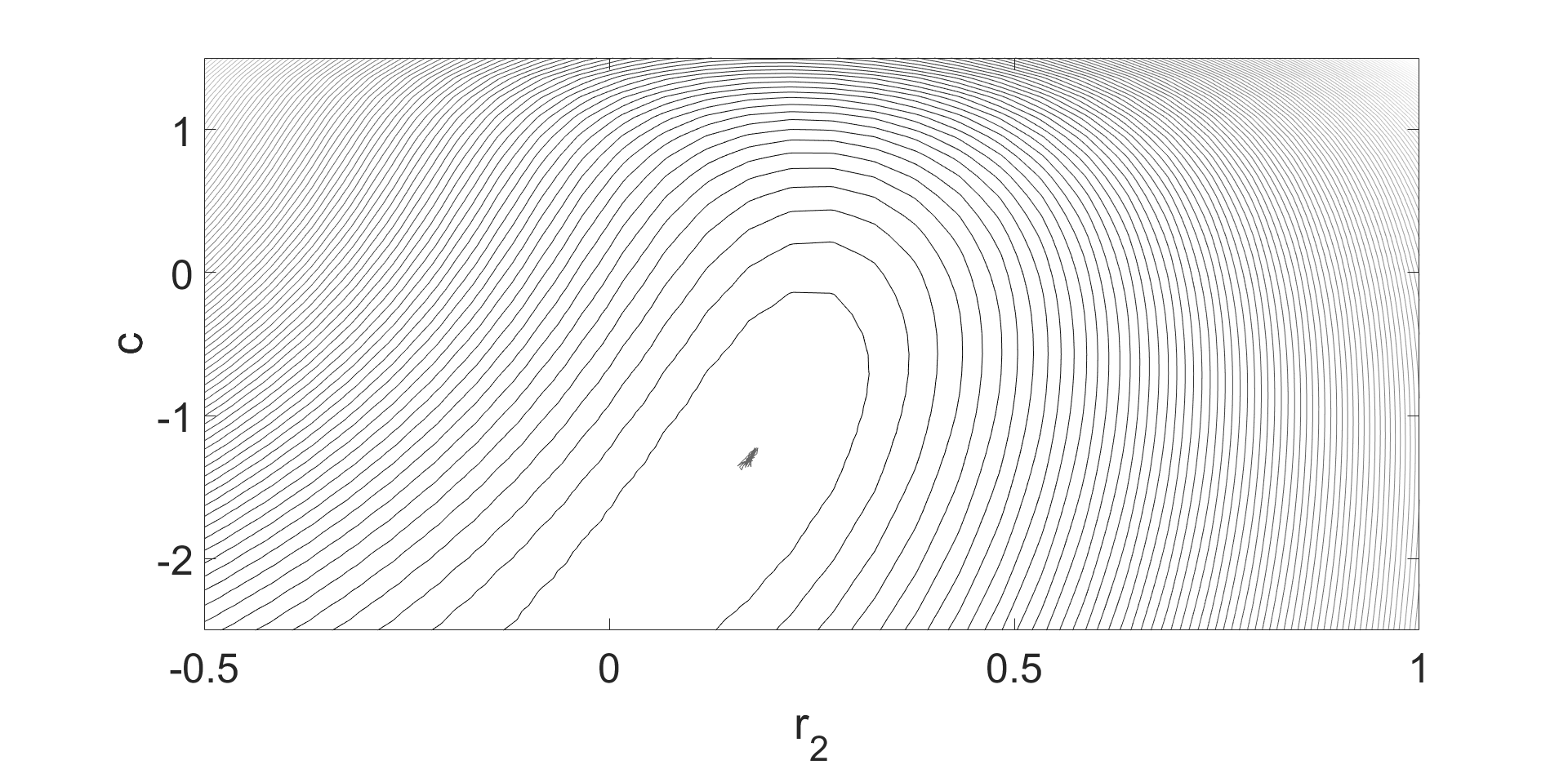}}
\subfigure[Residual sum-of-squares (S) for parameters $f$ and $c$ with $S: (1.2\times10^2, 1.0\times10^5)$]{\label{fig:2d}\includegraphics[width=0.75\textwidth]{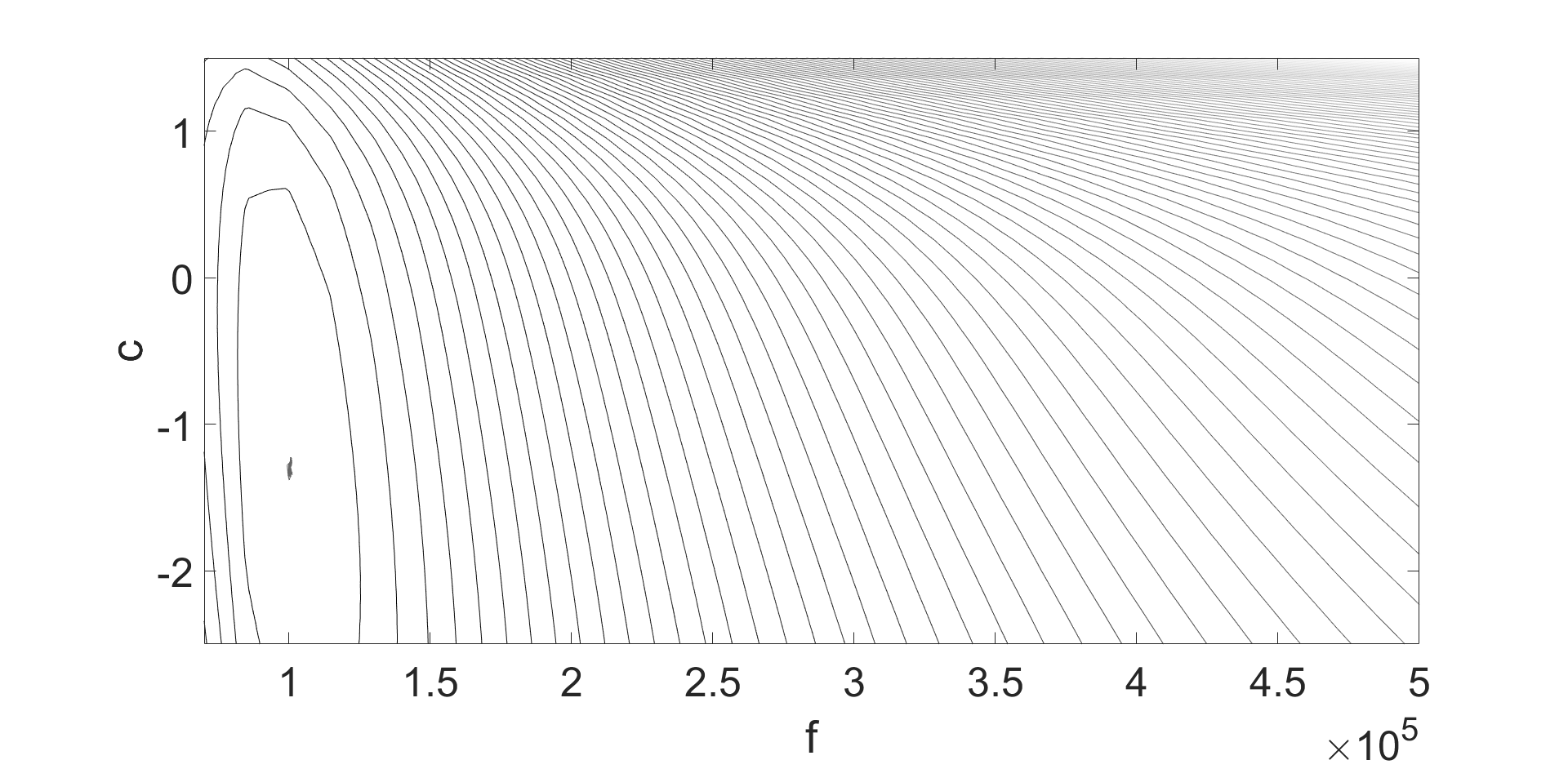}}
\caption{Contour plots of the objective function (residual sum-of-squares) in 2D, decreasing in the direction towards the inner slices for the 4D model corresponding to a control mouse. The black trajectories in the innermost slice indicate parameter (defined in the Simulations section) samples from the posterior distribution obtained using the Delayed Rejection Adaptive Metropolis algorithm. For graphical visibility, the innermost slice is not further resolved with additional contour lines.}
\label{fig:2}
\end{figure}

\begin{figure}[h!]
\centering
\subfigure[Residual sum-of-squares (S) for parameters $r_2$ and $\xi$ with $S: (1.0\times10^2, 1.7\times10^4)$]{\label{fig:3a}\includegraphics[width=0.75\textwidth]{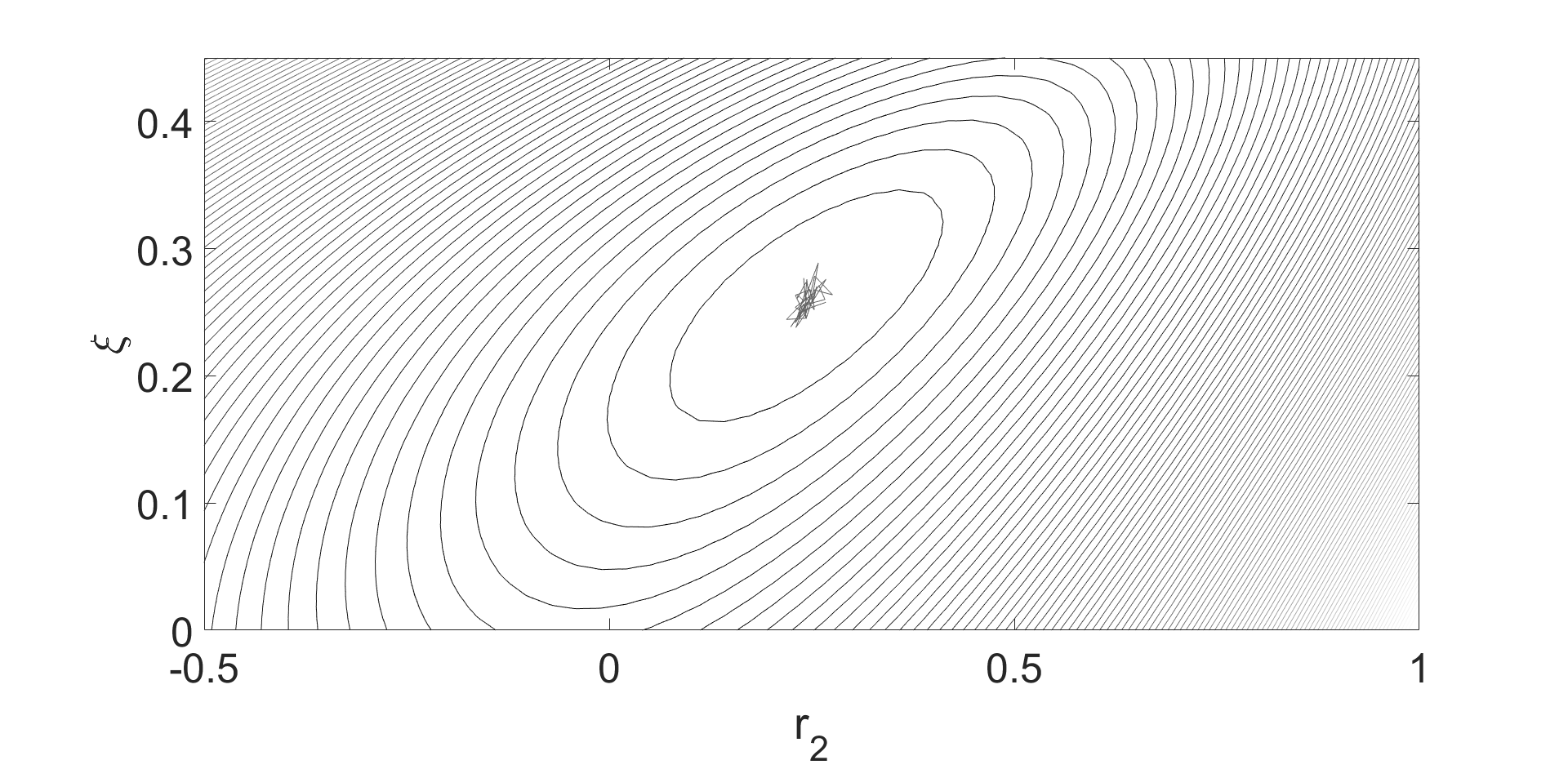}}
\subfigure[Residual sum-of-squares (S) for parameters $f$ and $\xi$ with $S: (1.0\times10^2, 1.6\times10^5)$]{\label{fig:3c}\includegraphics[width=0.75\textwidth]{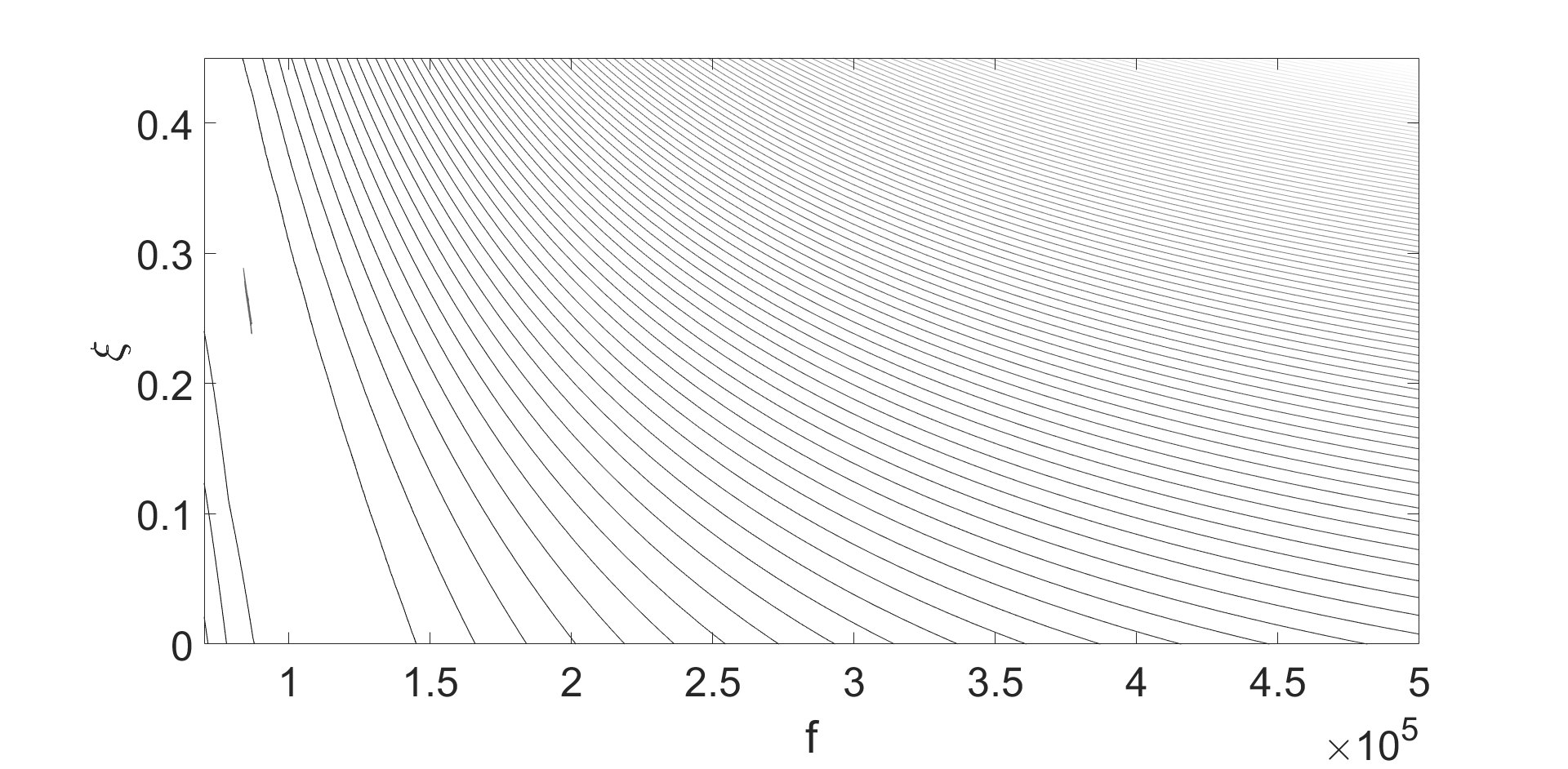}}
\subfigure[Residual sum-of-squares (S) for parameters $f$ and $r_2$ with $S: (1.1\times10^2, 1.1\times10^5)$]{\label{fig:3d}\includegraphics[width=0.75\textwidth]{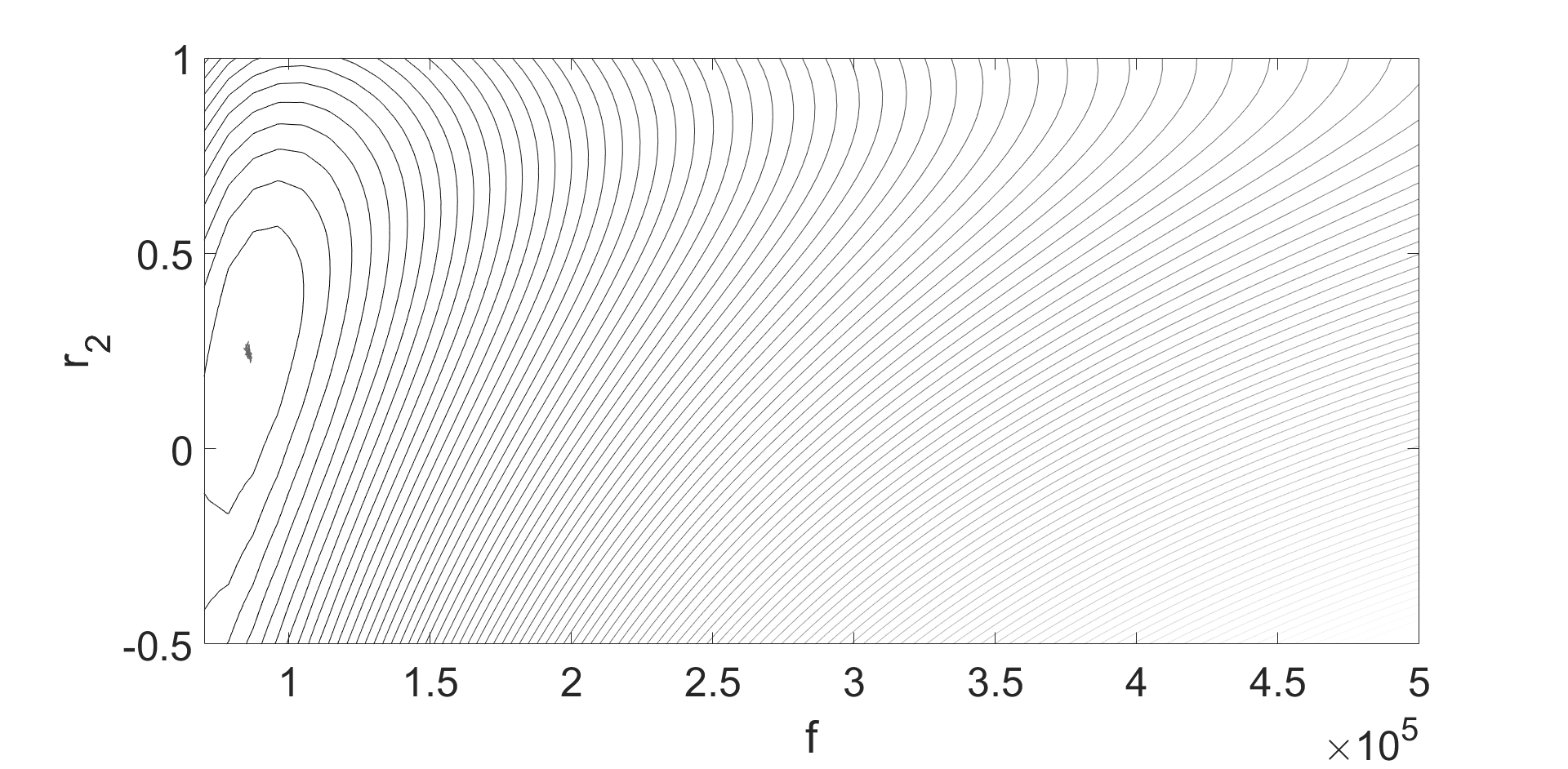}}
\caption{Contour plots of the objective function (residual sum-of-squares) in 2D, decreasing in the direction towards the inner slices for the 5D model corresponding to a control mouse. The black trajectories in the innermost slice indicate parameter (defined in the Simulations section) samples from the posterior distribution obtained using the Delayed Rejection Adaptive Metropolis algorithm. For graphical visibility, the innermost slice is not further resolved with additional contour lines.}
\label{fig:3}
\end{figure}

We start our analysis with an exploration of the objective function S ( equation (\ref{eq:5})), calculated for pressure in the MPA. We can visualize the landscape of S in 1D and 2D space (in 2D we vary two parameters at the time, and in 1D we vary one parameter at the time, while keeping all the others fixed to some value). We choose a representative mouse (the control mouse) and illustrate some of the 2D plots of S in Figure \ref{fig:2} (4D model: no tapering) and Figure \ref{fig:3} (5D model: tapering).

\begin{figure}[h!]
\centering
\subfigure[Pressure obtained from optimization (dashed line) plotted alongside the measured pressure (solid line) for the 5D model corresponding to a control mouse, $S = 104.5$]{\label{fig:4a}\includegraphics[width=0.49\textwidth]{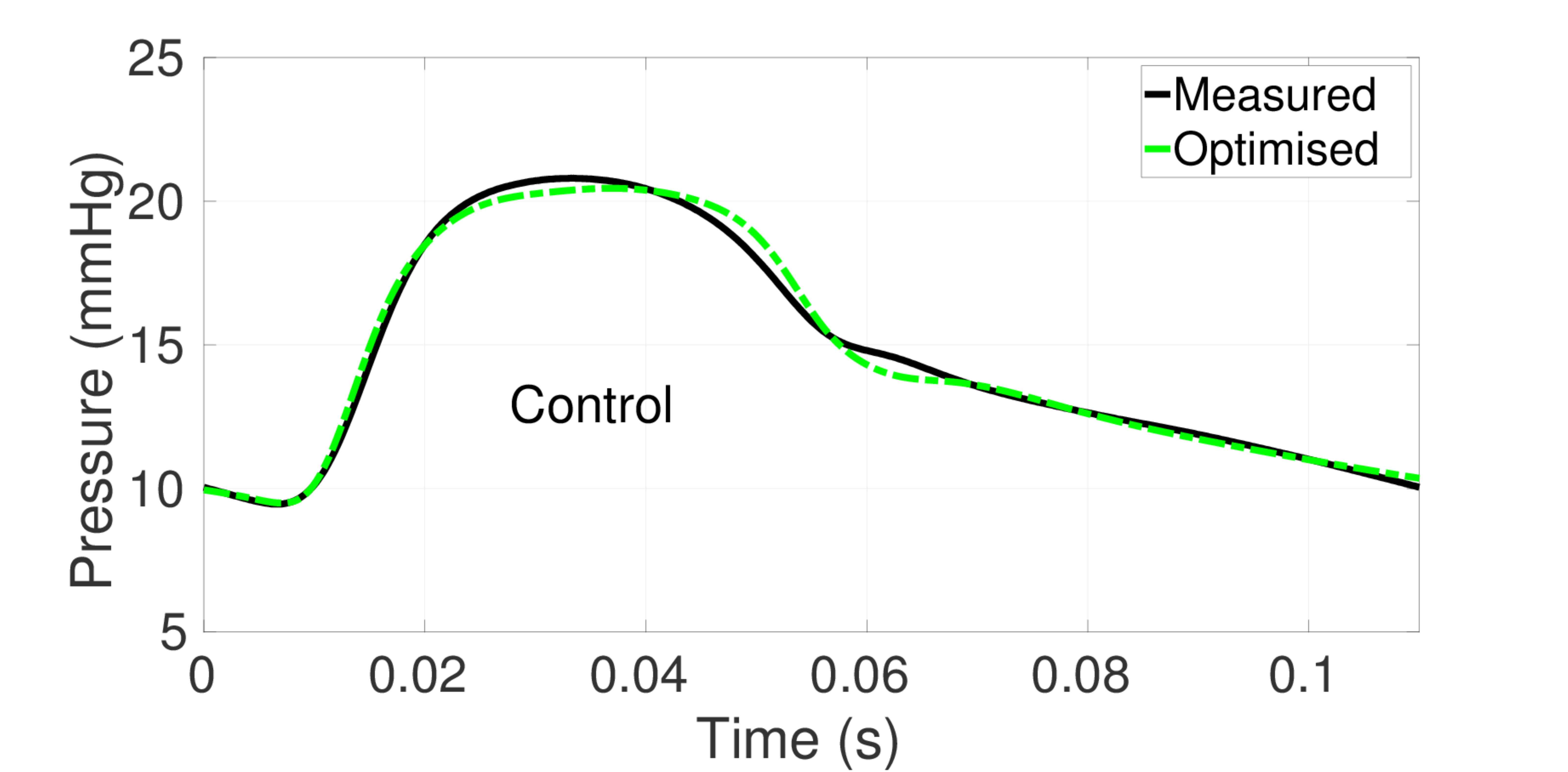}}
\subfigure[Pressure obtained from optimization (dashed line) plotted alongside the measured pressure (solid line) for the 5D model corresponding to a hypoxic mouse, $S = 80$]{\label{fig:4b}\includegraphics[width=0.49\textwidth]{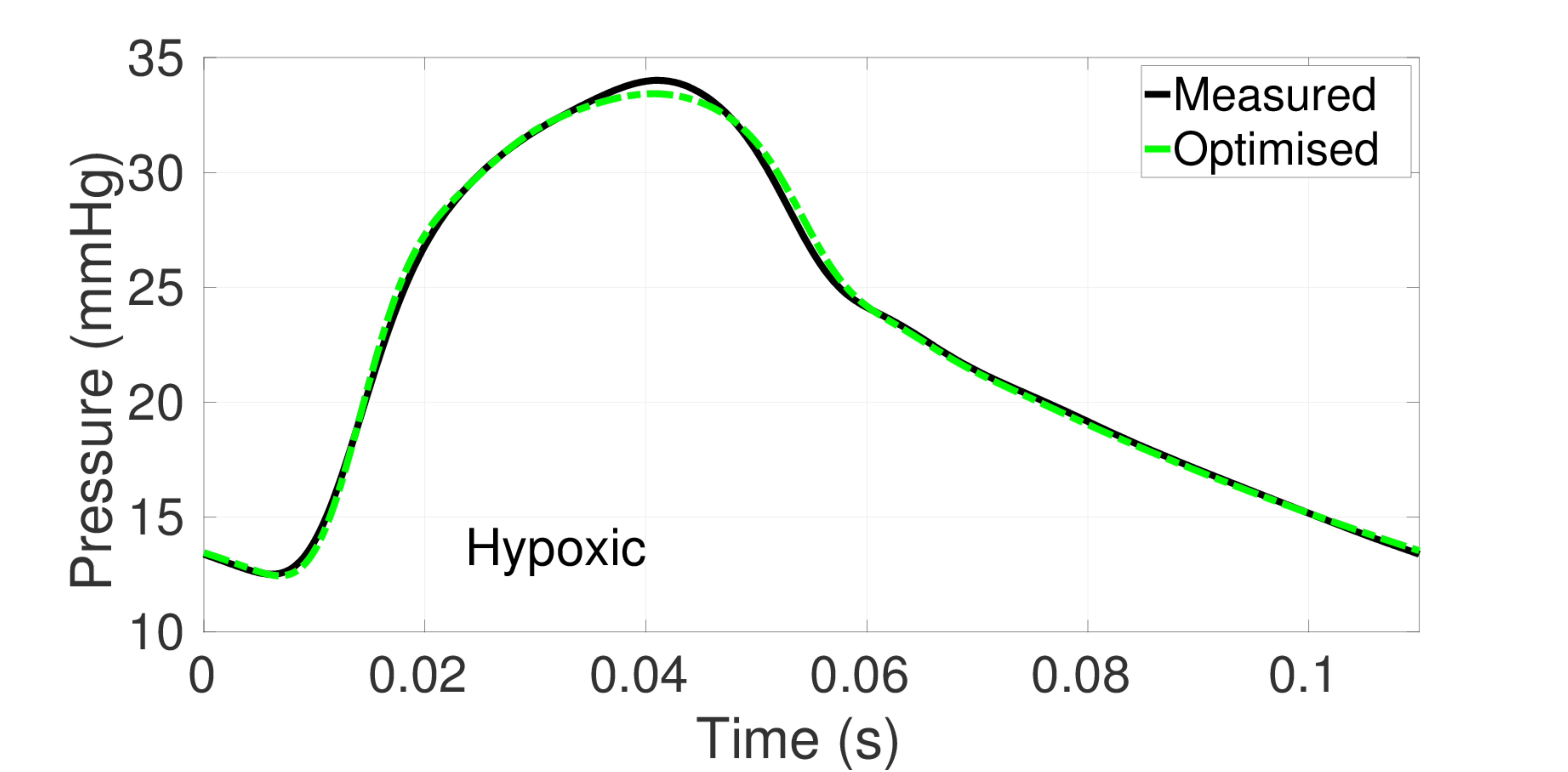}}
\caption{Comparison between measured pressure (solid line) and pressure obtained from optimization (dashed line) for the 5D model for a control and a hypoxic mouse.}
\label{fig:4}
\end{figure}

S exhibits unimodality and it is highly skewed in 2D, and this is the case for both the control and the hypoxic mouse tested, for both models. In some of the figures, there is a sudden jump from low S values to high S values ($10^{10}$). These very high S value region marks the parameter domain that is invalid under the mathematical model. This shows that the biological parameter bounds are valid in an univariate sense, but this domain gets more restricted in higher dimensions, with some parameter combinations that produce no solutions to the PDEs (\ref{eq:1}) - (\ref{eq:2}), due to being outside the domain of convergence of the numerical scheme. The unimodality in S motivates applying the nonlinear constraint optimization to our problem to find the set of parameters that minimise S. We do so for all of the parameters in the model (see equation (\ref{eq:12})), for both mice, and to ensure good coverage of the multidimensional space, we generate 20 initial values from a Sobol sequence. Our results reveal that regardless of the starting value, we manage to recover the same parameter values.
control mouse: $f = 85529; r_1 = 1.96; r_2 = 0.25; c = -2.03; \xi = 0.26$, and $S_{\textrm{final}} = 104.5$; 
hypoxic mouse: $f = 228383; r_1 = 0.59; r_2 = 0.14; c = -0.22; \xi = 0.09$, and $S_{\textrm{final}} = 80$,
and we consider these to be the maximum likelihood estimates. We can use these estimates as input into the PDEs to predict the pressure signal, and compare this against the measured pressure (Figure \ref{fig:4}), and we do so for both mice. 

\begin{figure}[h!]
\centering
\subfigure[Pressure obtained from optimization (dashed line) plotted alongside the measured pressure (solid line) for the 4D model corresponding to a control mouse, $S = 126.5$]{\label{fig:5a}\includegraphics[width=0.49\textwidth]{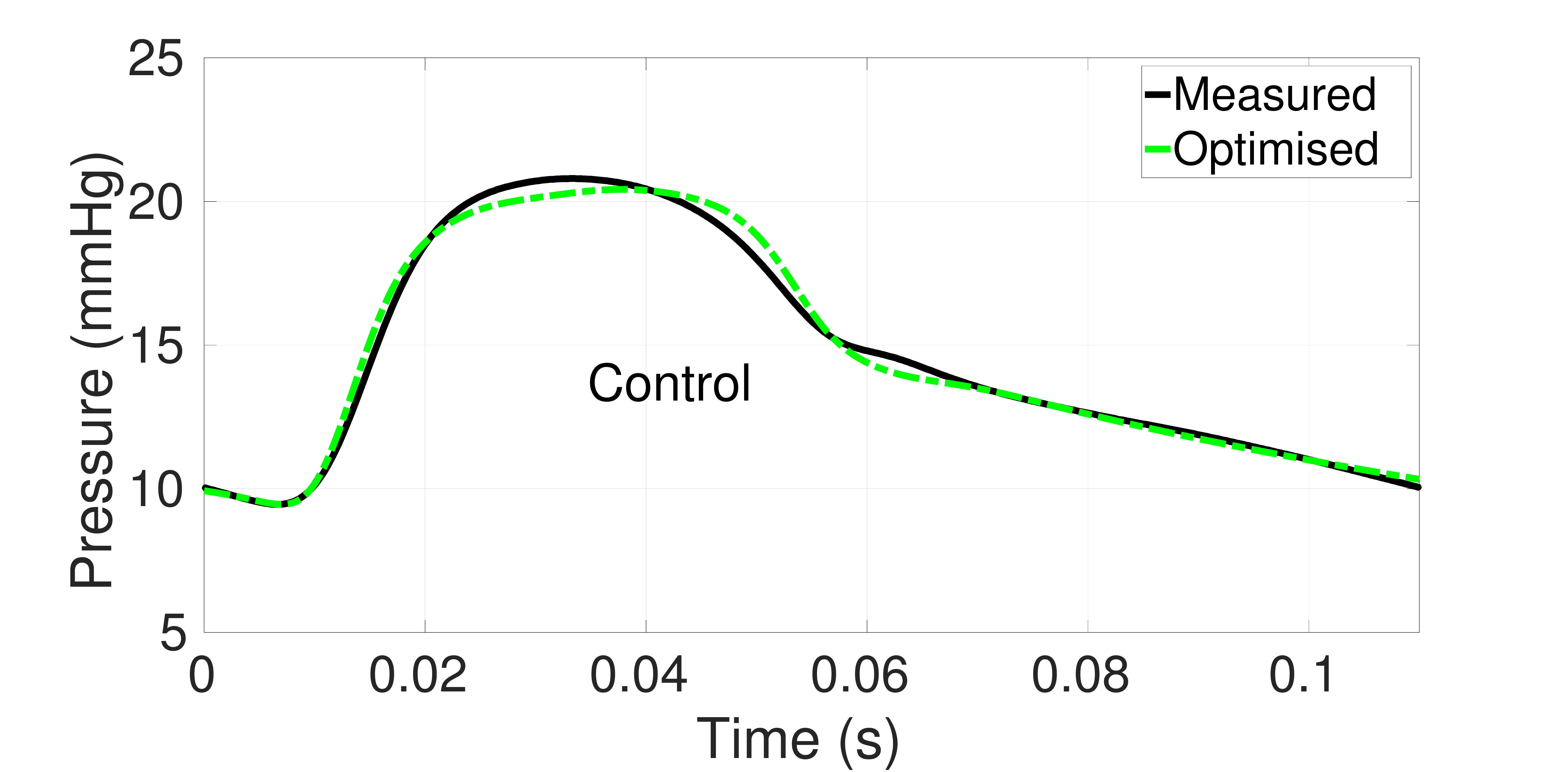}}
\subfigure[Pressure obtained from optimization (dashed line) plotted alongside the measured pressure (solid line) for the 4D model corresponding to a hypoxic mouse, $S = 80.3$]{\label{fig:5b}\includegraphics[width=0.49\textwidth]{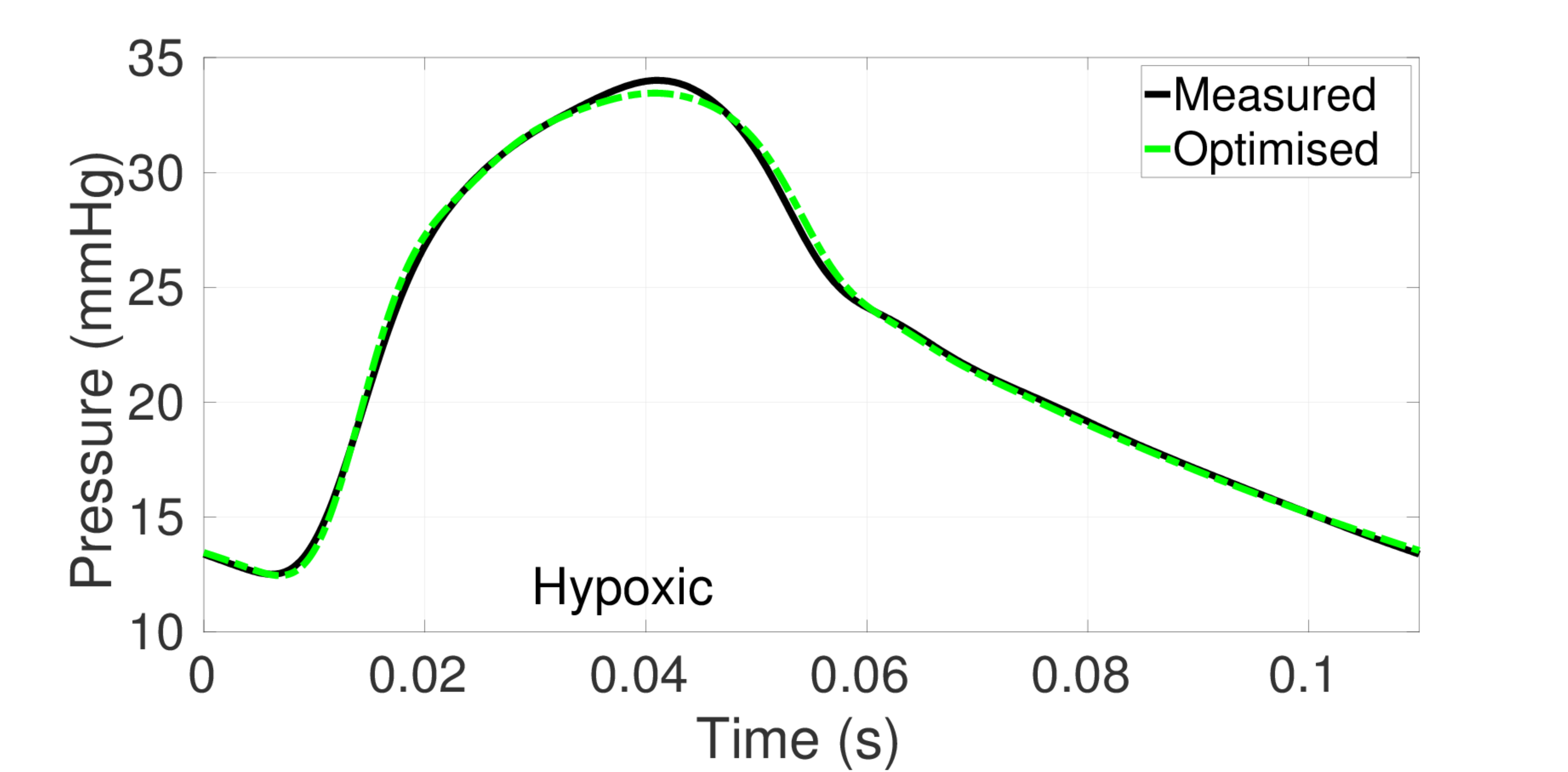}}
\caption{Comparison between measured pressure (solid line) and pressure obtained from optimization (dashed line) for the 4D model for a control and a  hypoxic mouse.}
\label{fig:5}
\end{figure}

Figure \ref{fig:4} shows the optimised pressure waveform for the 5D problem plotted alongside the measured pressure data for both mice. Panel (a) shows that in the case of the healthy mouse, the simulated pressure closely follows the measured pressure except near the peak, where an offset is registered. The overall model prediction appears to be better for the hypoxic mouse (panel (b)), which is also reinforced by the lower S score.

Next, we set $\xi = 0$ and repeat the analysis for a parameter set that excludes $\xi$. Our findings indicate that for the Control mouse, the fit is poorer if $\xi$ is not included, but the fit is very similar for the Hypoxic mouse.
Control mouse: $f = 100398; r_1 = 1.69; r_2 = 0.17; c = -1.30$, and $S_{\textrm{final}} = 126.5$;
Hypoxic mouse: $f = 243645; r_1 = 0.52; r_2 = 0.13; c =  -0.24$, and $S_{\textrm{final}} = 80.3$.
See Figure \ref{fig:5} for a visual inspection of the fit.

However, we cannot decide on whether to include $\xi$ in our model without a formal model selection analysis, which will follow later in the paper.

\subsection{MCMC Results}
Next, we aim to quantify the uncertainty around our parameter estimates, and to do so, we implement the DRAM algorithm, the AM algorithm and the DR algorithm to our problem. This is done for each mouse for the two models (with and without the taper factor, $\xi$).

\subsubsection{AM Results}
In the AM algorithm there are a couple of parameters that need to be set prior to the commencement of the algorithm. One such parameter is $s_d$ (see (\ref{eq:8})), which we take to be the recommended value in literature, $2.4^2/d$. Another such parameter is the length of the adaptation interval, $t_{\textrm{ad}}$ (see (\ref{eq:8})), which in practice is chosen to ensure mixing while having enough points that we use for updating the chain covariance matrix. The recommended $t_{\textrm{ad}}$ is 100 in literature, however we have found that for our problem, a much larger $t_{\textrm{ad}}$ was needed. We experimented with values of 100, 500 and 1000. Our findings indicate that $t_{\textrm{ad}} = 1000$ gives higher acceptance rates by up to $5\%$ than $t_{\textrm{ad}} = 500$. The acceptance rate stays between $19\% - 23\%$, and the mixing appears to be better than for $t_{\textrm{ad}} = 100$, so we have chosen $t_{\textrm{ad}} = 1000$.
In Figures \ref{fig:6} to \ref{fig:11} (in the text), and \ref{fig:10} and \ref{fig:12} (in the Appendix) we show traceplots, marginal posterior densities and scatterplots of parameters sampled during the AM simulation for the two mice, for both models and Table \ref{table:1} provides posterior summaries for the parameters, and we notice that the posterior means are very similar to the optimised values from the Optimization Results section.

For the control mouse, for the two models (Figures \ref{fig:6} - \ref{fig:9} in the text and \ref{fig:10} in the Appendix), we can see that the chains fluctuate around the parameter values obtained from optimisation, suggesting that the chain has reached the highest posterior density region, and that the optimization might have found the global optimum.

There are no signs of multimodality in the objective function, which was reinforced by the convergence results we obtained through optimization. In addition, the S values are steady around the initial, lowest S value and the $\sigma^2$ chain does not indicate non-convergence (see Figure \ref{fig:7}). Also, the marginal posterior densities for some parameters have long tails (Figure \ref{fig:8}), which reinforces the skewness evident in the 2D S plots (Figures \ref{fig:2} and \ref{fig:3}). The scatterplots (Figures \ref{fig:9}) reveal strong correlations between some of the parameters, which is a reason why the algorithm needs longer runs to converge. For example, the stiffness $f$ is highly correlated to the tapering factor, $\xi$, as the wall stiffness increases, the vessels will become less tapered (top and bottom vessel radii get closer together), which is what we would expect (see (\ref{eq:14})). Another example is the high correlation between stiffness, $f$ and compliance adjustment parameter, $c$, i.e. as the stiffness of the vessel wall increases, the compliance adjustment increases, and so the nominal compliance or elasticity decreases, which is expected (see (\ref{eq:13})).

\begin{figure}[h!]
\centering
\includegraphics[scale=0.25]{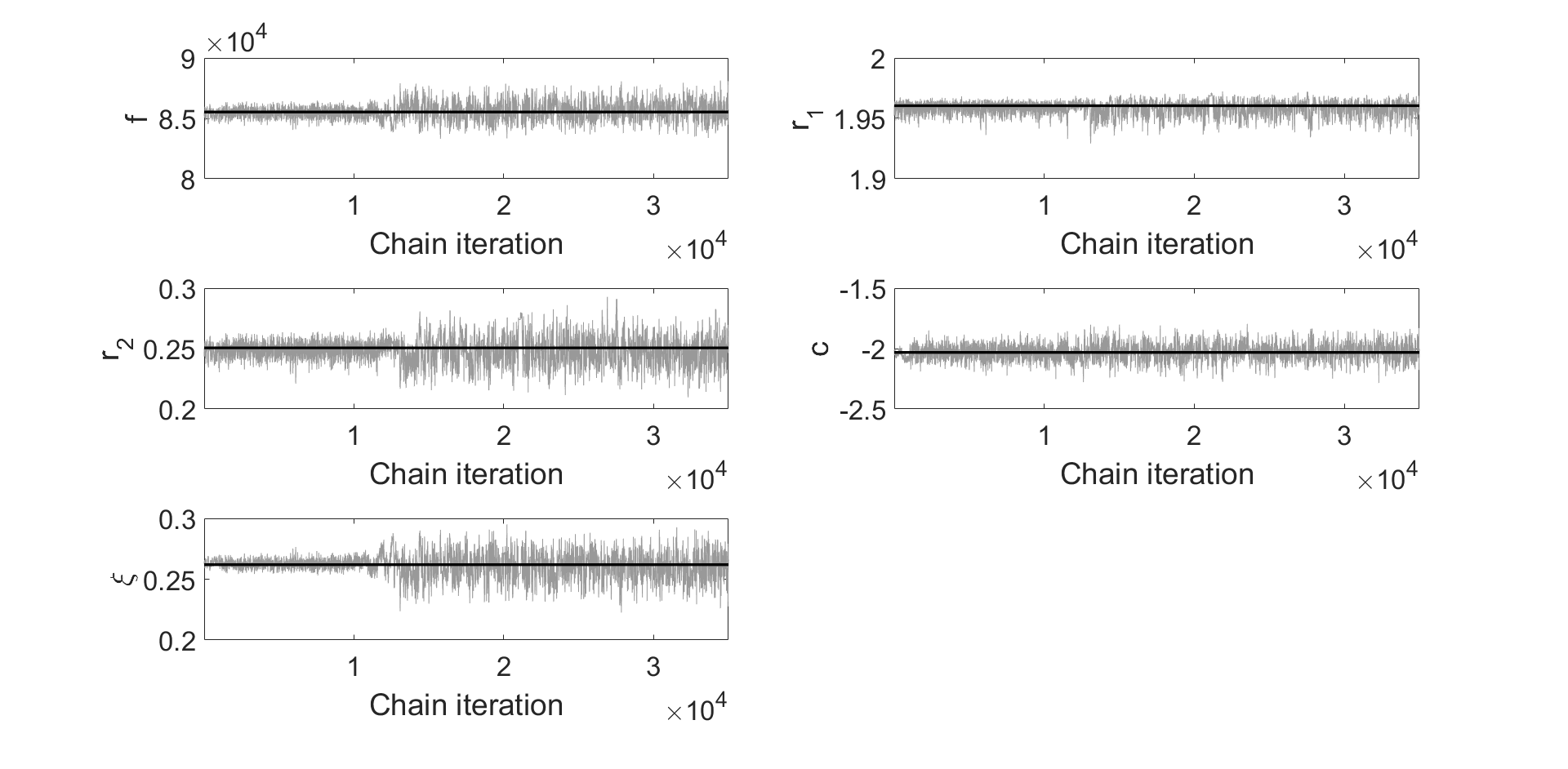}
\caption{Markov chains traceplots for the parameters (defined in the Simulations section) obtained using the Adaptive Metropolis algorithm for the 5D model corresponding to a control mouse. Starting values for the algorithm are the optimised values and are superimposed in black horizontal lines. Acceptance rate is $22\%$.}
\label{fig:6}
\end{figure}

\begin{figure}[h!]
\centering
\includegraphics[scale=0.25]{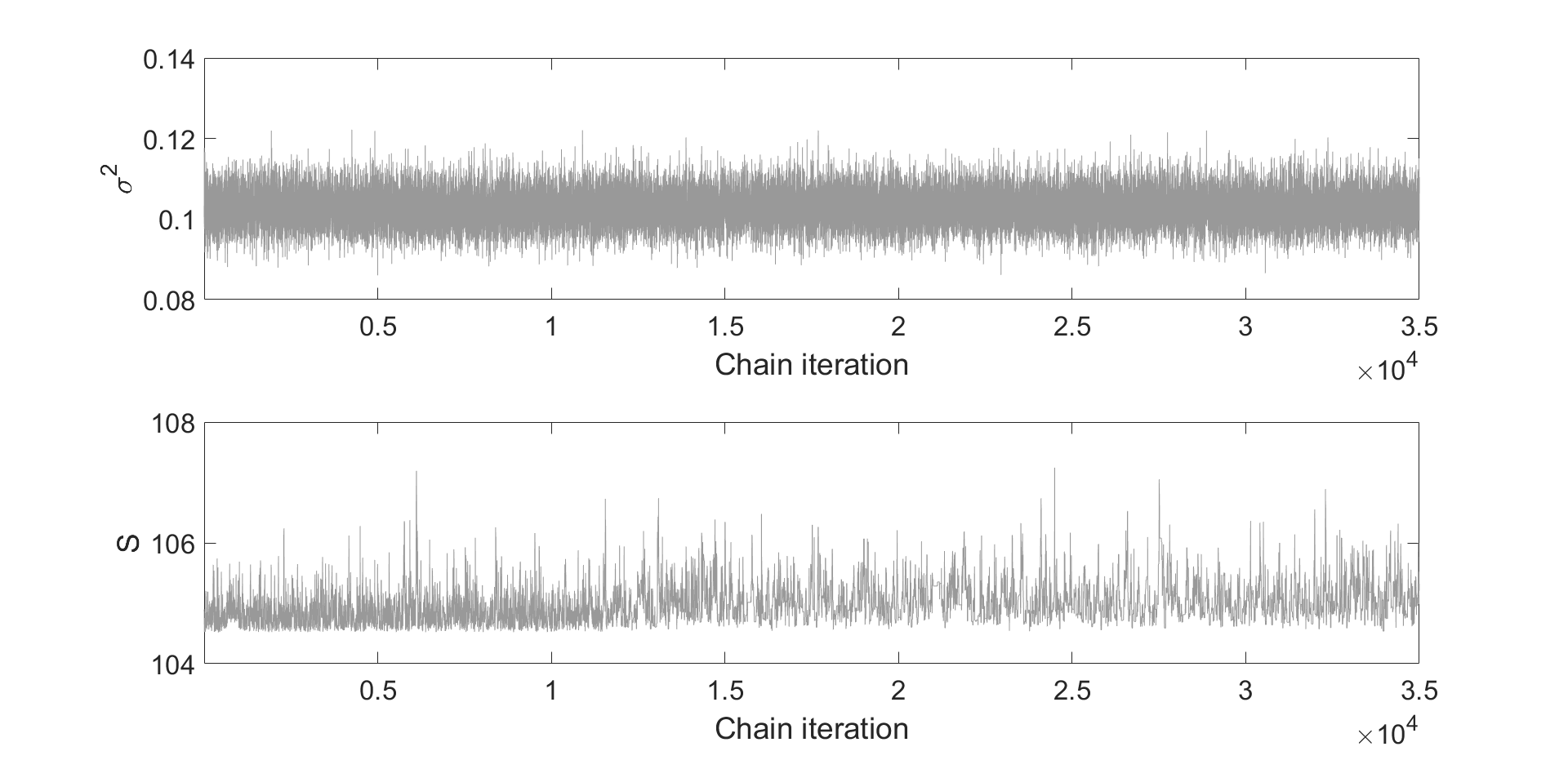}
\caption{Markov chains traceplots for the error variance $\sigma^2$, and the residual-sum-of-squares $S$ obtained using the Adaptive Metropolis algorithm for the 5D model corresponding to a control mouse.}
\label{fig:7}
\end{figure}

\begin{figure}[h!]
\centering
\includegraphics[scale=0.25]{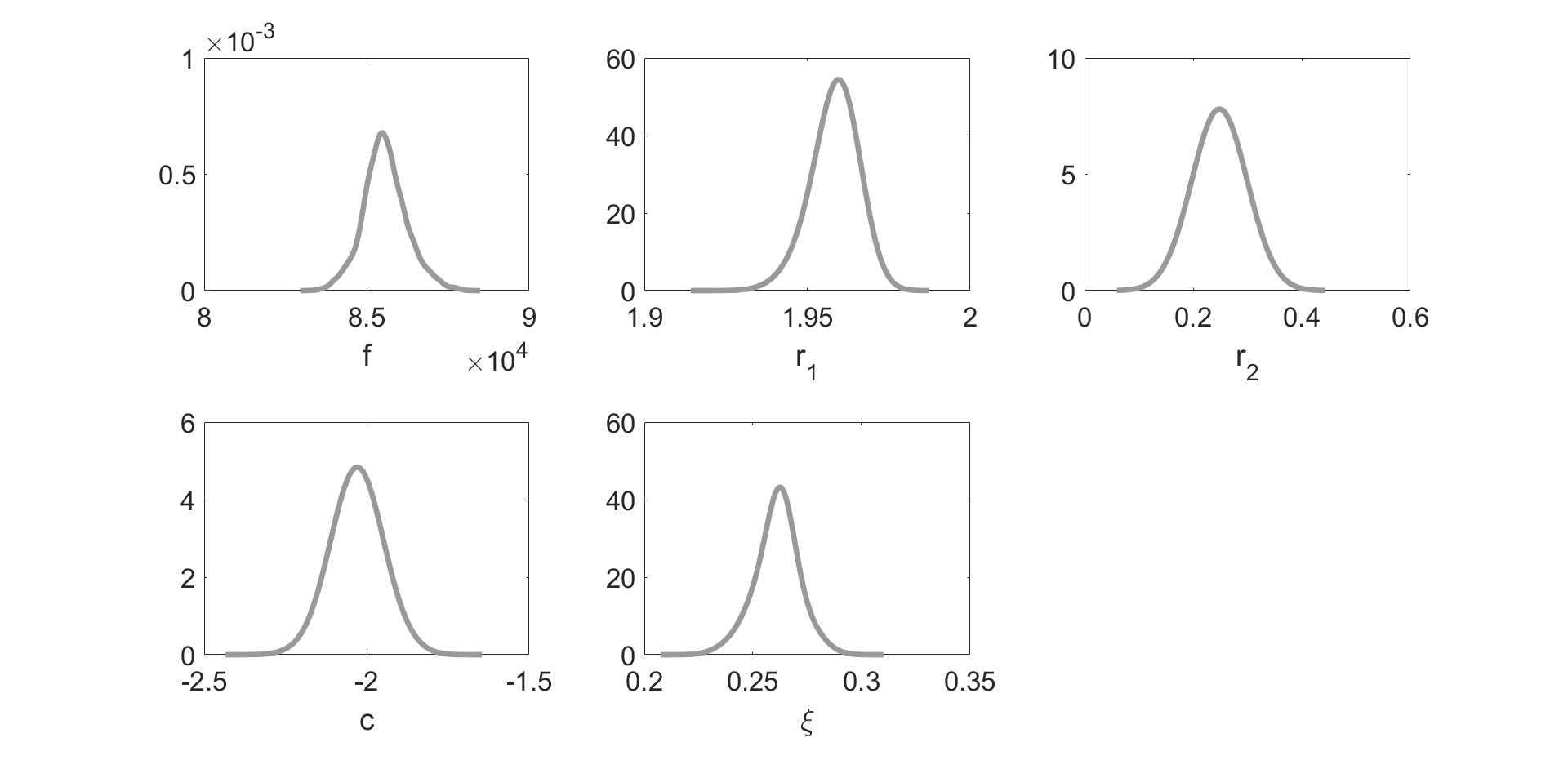}
\caption{Marginal posterior densities of parameters (defined in the Simulations section) inferred using the Adaptive Metropolis algorithm for the 5D model corresponding to a control mouse. Plots were obtained using a kernel density estimator with customised bandwidth.}
\label{fig:8}
\end{figure}

\begin{figure}[h!]
\centering
\includegraphics[scale=0.25]{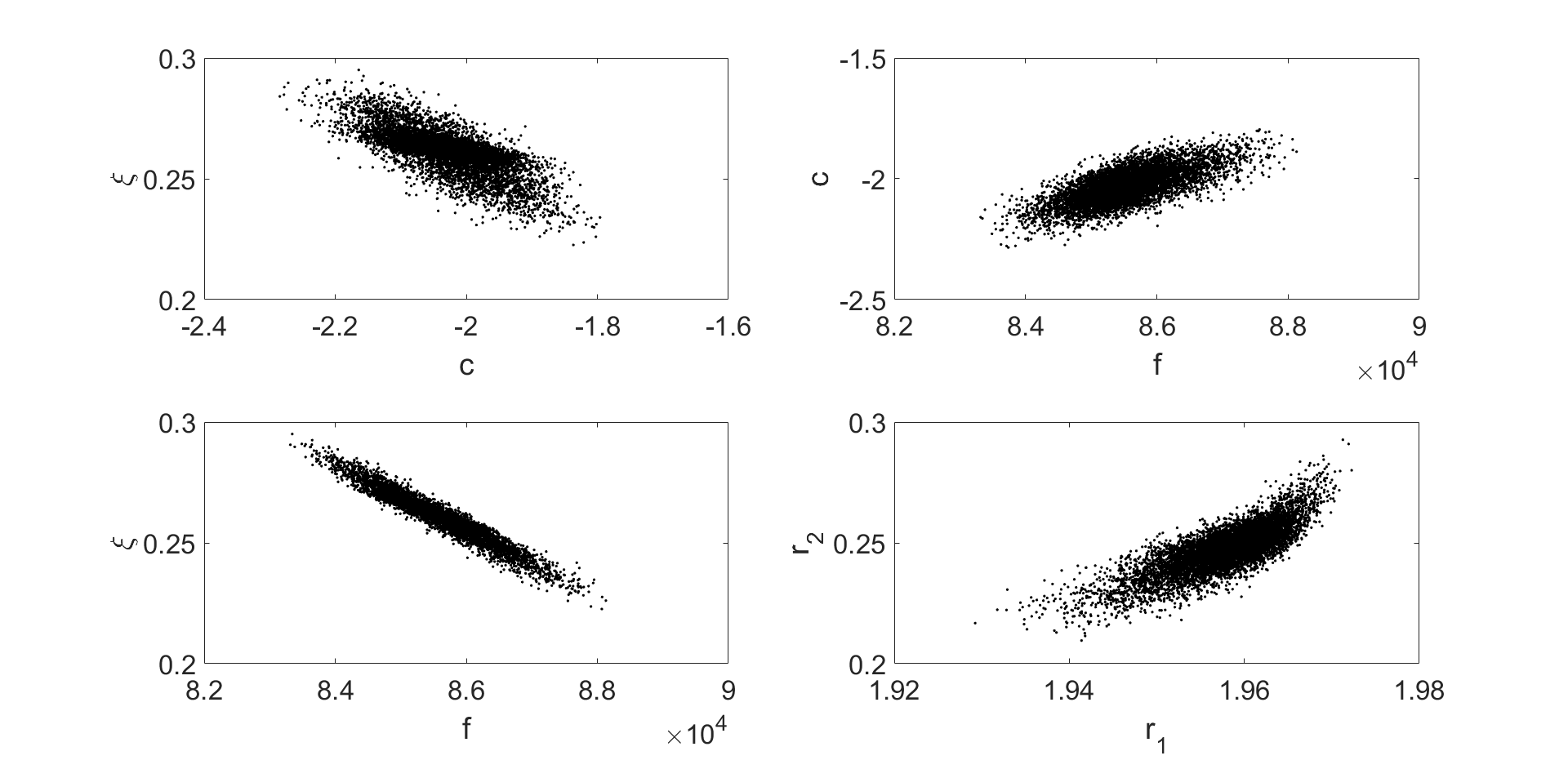}
\caption{Scatterplots of parameter (defined in the Simulations section) chains obtained using the Adaptive Metropolis algorithm for the 5D model corresponding to a control mouse.}
\label{fig:9}
\end{figure}

Regarding the hypoxic mouse, for the 5D model, we note a more pronounced skewness in the distribution for two of the parameters: $f$ and $\xi$ (Figure \ref{fig:11}), which are strongly correlated. We also notice that the biological range for $\xi$ is overly conservative, and a wider range should be used. In addition, the posterior mean appears to vary slightly from the maximum likelihood estimates for some of the parameters, but the two are more similar in the 4D model for the hypoxic mouse (Figure \ref{fig:12} in the Appendix).

\begin{figure}[h!]
\centering
\includegraphics[scale=0.25]{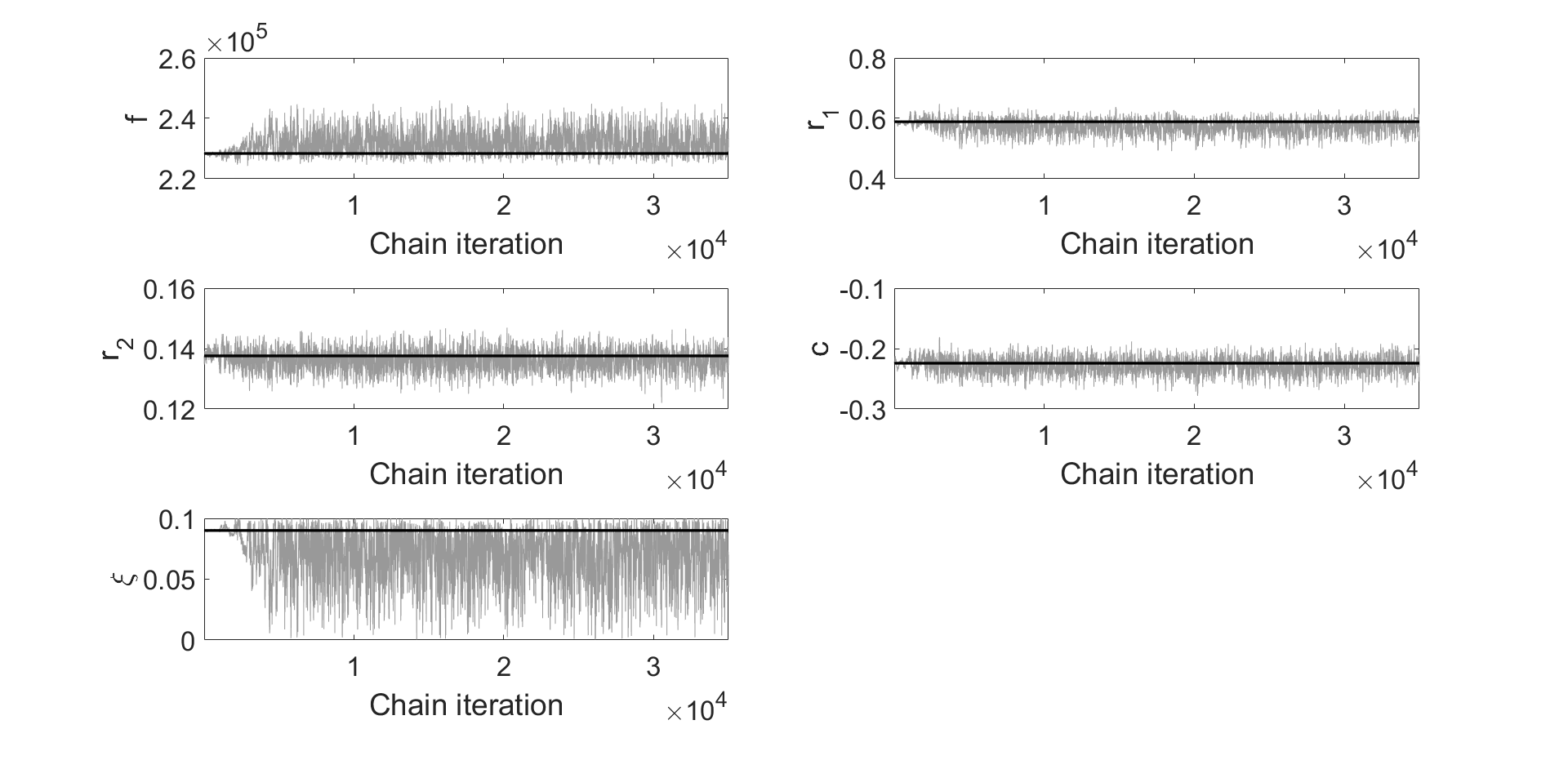}
\caption{Markov chains traceplots for the parameters (defined in the Simulations section) obtained using the Adaptive Metropolis algorithm for the 5D model corresponding to a hypoxic mouse. Starting values for the algorithm are the optimised values and are superimposed in black horizontal lines. Acceptance rate is $23\%$.}
\label{fig:11}
\end{figure}

\clearpage
\subsubsection{DR Results}

\begin{figure}[h!]
\centering
\includegraphics[scale=0.3]{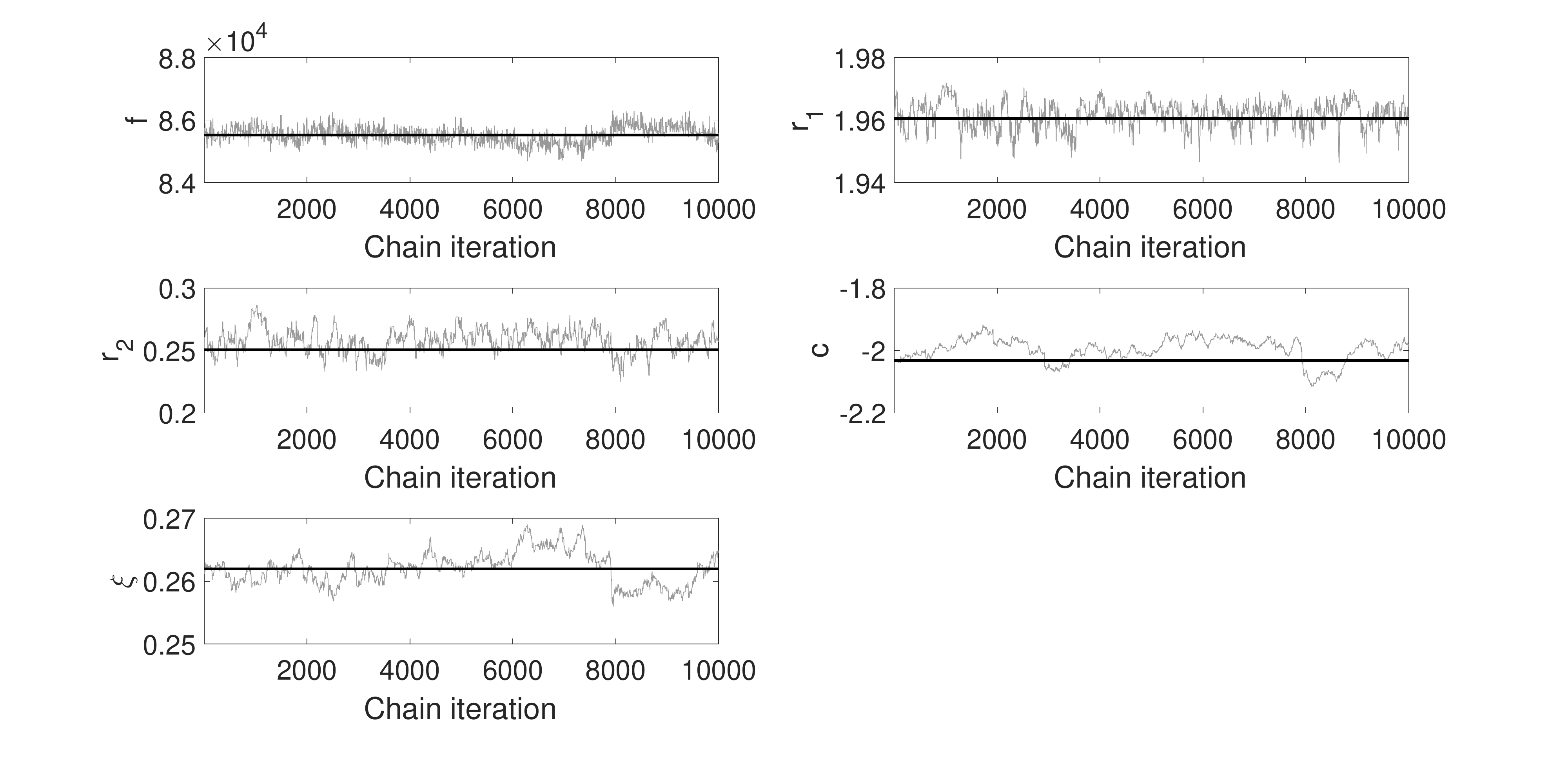}
\caption{Markov chains traceplots for the parameters (defined in the Simulations section) obtained using the Delayed Rejection algorithm for the 5D model corresponding to a control mouse. Starting values for the algorithm are the optimised values and are superimposed in black horizontal lines. Acceptance rate is $22\%$.}
\label{fig:13}
\end{figure}

Figure \ref{fig:13} shows 10000 iterations of a chain produced using the DR algorithm for the control mouse, 5D model. Despite the acceptance rate which is close to the optimal target acceptance rate given in literature ($23\%$), the plots reveal non-convergence of the chains, and the parameter space is explored slowly, leading to bad mixing. DR seems to be affected by the strong correlations between the parameters. Therefore, the chains would have to be run for a very long time, implying that the AM algorithm is essential in reaching convergence in a reasonable time frame.

\subsubsection{DRAM Results}
Lastly, we implement the DRAM algorithm starting from optimised values for every mouse and model type. We find that the estimates are very similar to the estimates obtained using the AM algorithm (see Table \ref{table:1}), which is expected. The algorithm used to sample draws is irrelevant if the samples are approximately drawn from the posterior distribution; the algorithm only influences the convergence of the chains. This is supported by the traceplots showing these samples (Figures \ref{fig:14} and \ref{fig:15} in the text and \ref{fig:16}, \ref{fig:17}, \ref{fig:18} in the Appendix), which do not indicate non-convergence, as the chains appear to be exploring the space around the optimum values well and are steady around these values. We find that the DRAM algorithm has increased the acceptance rate compared to the AM algorithm (from $20\%$ up to $40\%$). This is not surprising, considering the extra DR step in the DRAM algorithm, which proposes a second point at every iteration of the algorithm. This clearly increases the acceptance rate, and thus, fewer iterations are needed to reach convergence when compared to AM. This, however, does not imply less computational time, since in every DRAM iteration, when a second point is proposed, the PDEs have to be solved twice. Therefore, AM and DRAM are computationally equivalent in our case, since double the number of iterations were needed for AM to decrease the correlation between chain samples. The inspection of the convergence diagnostics (discussed in the MCMC Convergence Diagnostics section below), as well as the improvement in the acceptance rate has made us favour DRAM over AM.

\begin{figure}[h!]
\centering
\includegraphics[scale=0.25]{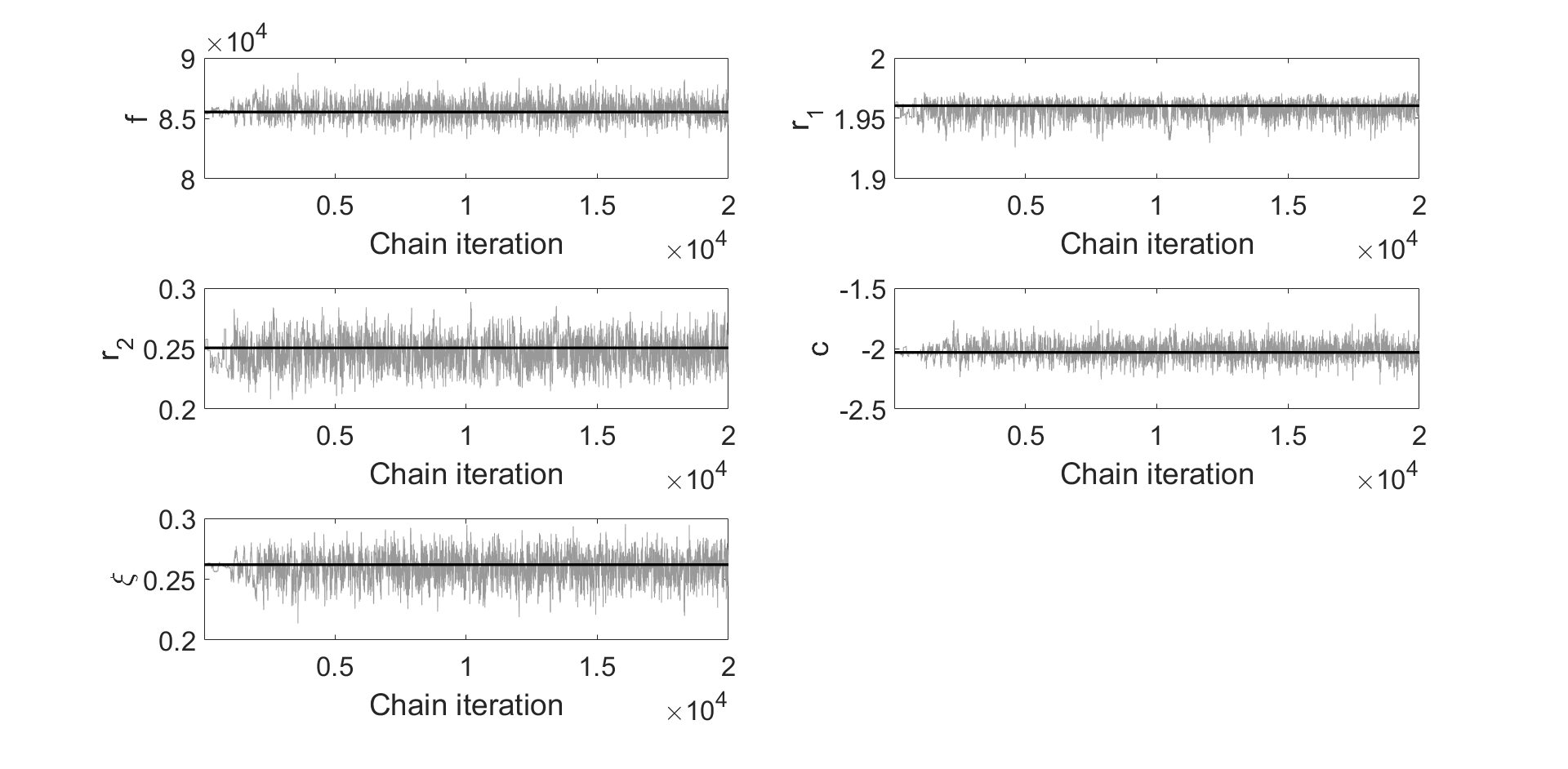}
\caption{Markov chains traceplots for the parameters (defined in the Simulations section) obtained using the Delayed Rejection Adaptive Metropolis algorithm for the 5D model corresponding to a control mouse. Starting values for the algorithm are the optimised values and are superimposed in black horizontal lines. Acceptance rate is $39\%$.}
\label{fig:14}
\end{figure}

\begin{figure}[h!]
\centering
\includegraphics[scale=0.25]{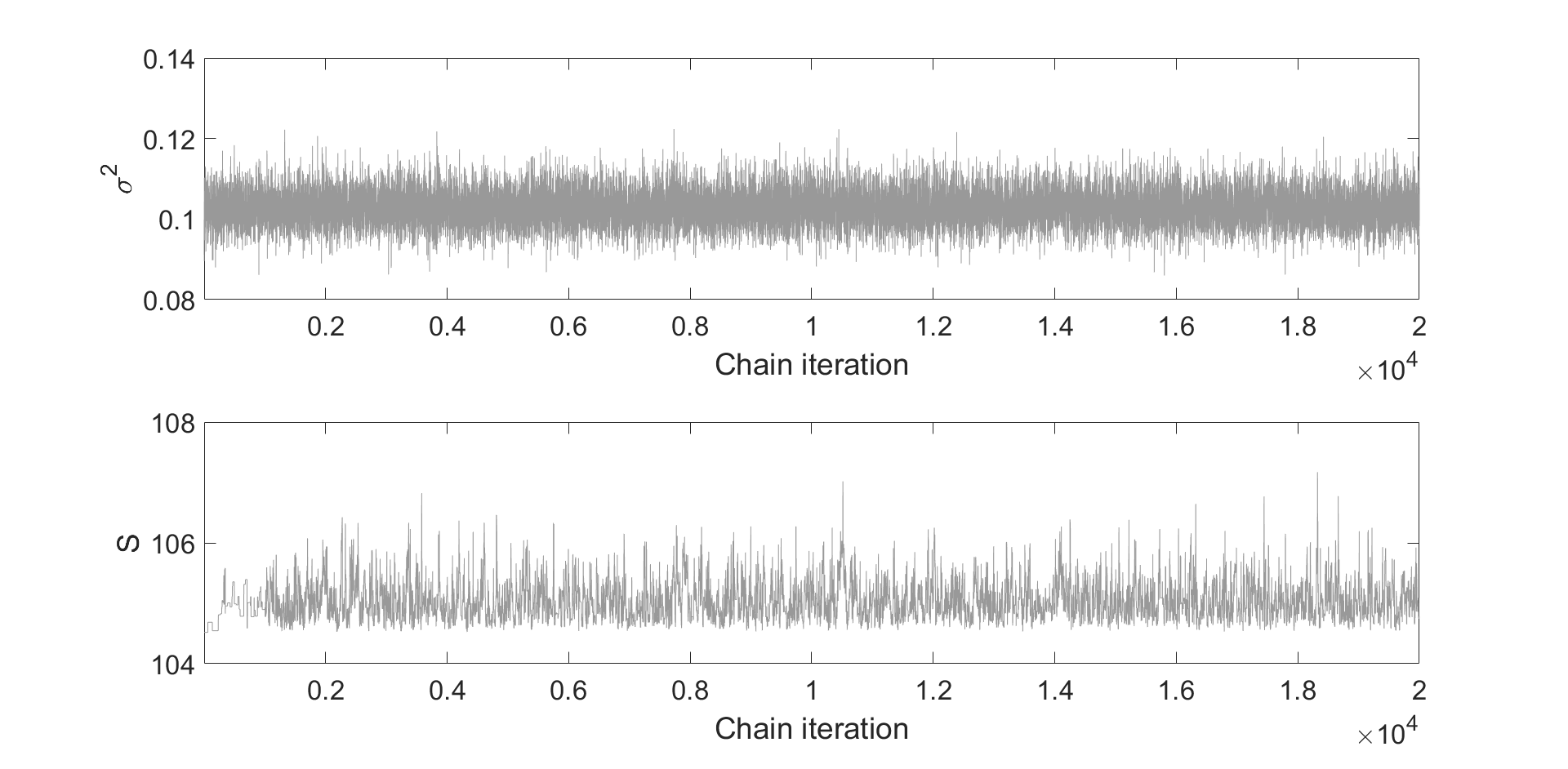}
\caption{Markov chains traceplots for the error variance $\sigma^2$, and the residual-sum-of-squares $S$ obtained using the Delayed Rejection Adaptive Metropolis algorithm for the 5D model corresponding to a control mouse.}
\label{fig:15}
\end{figure}

\begin{figure}[h!]
\centering
\includegraphics[scale=0.25]{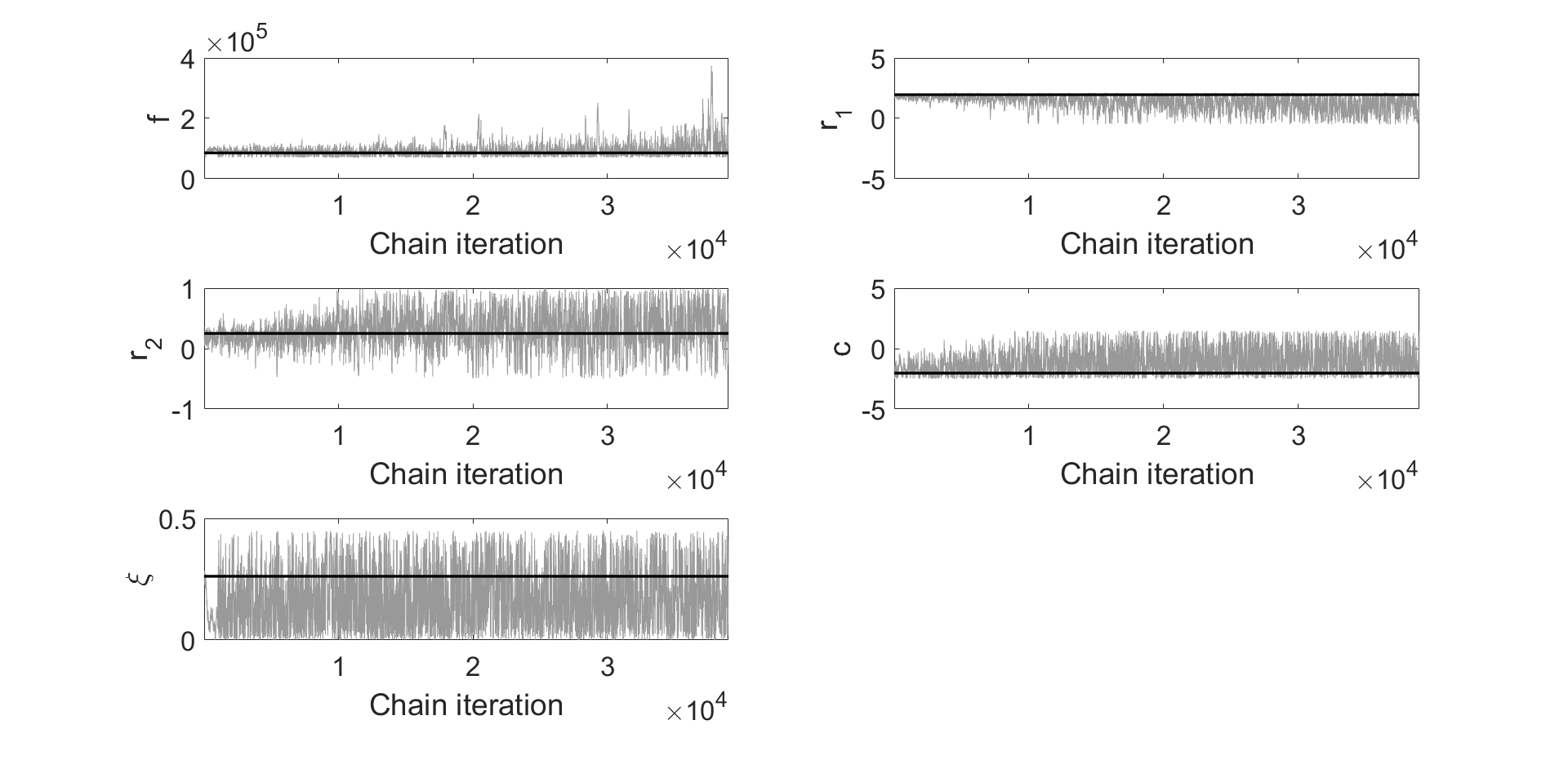}
\caption{Markov chains traceplots for the parameters (defined in the Simulations section) obtained using the original Delayed Rejection Adaptive Metropolis algorithm in \cite{UncertaintySmith} without the correction term discussed in this paper, for the 5D model corresponding to a control mouse. Starting values for the algorithm are the optimised values and are superimposed in black horizontal lines. Acceptance rate is $30\%$.}
\label{fig:19}
\end{figure}

\begin{figure}[h!]
\centering
\includegraphics[scale=0.25]{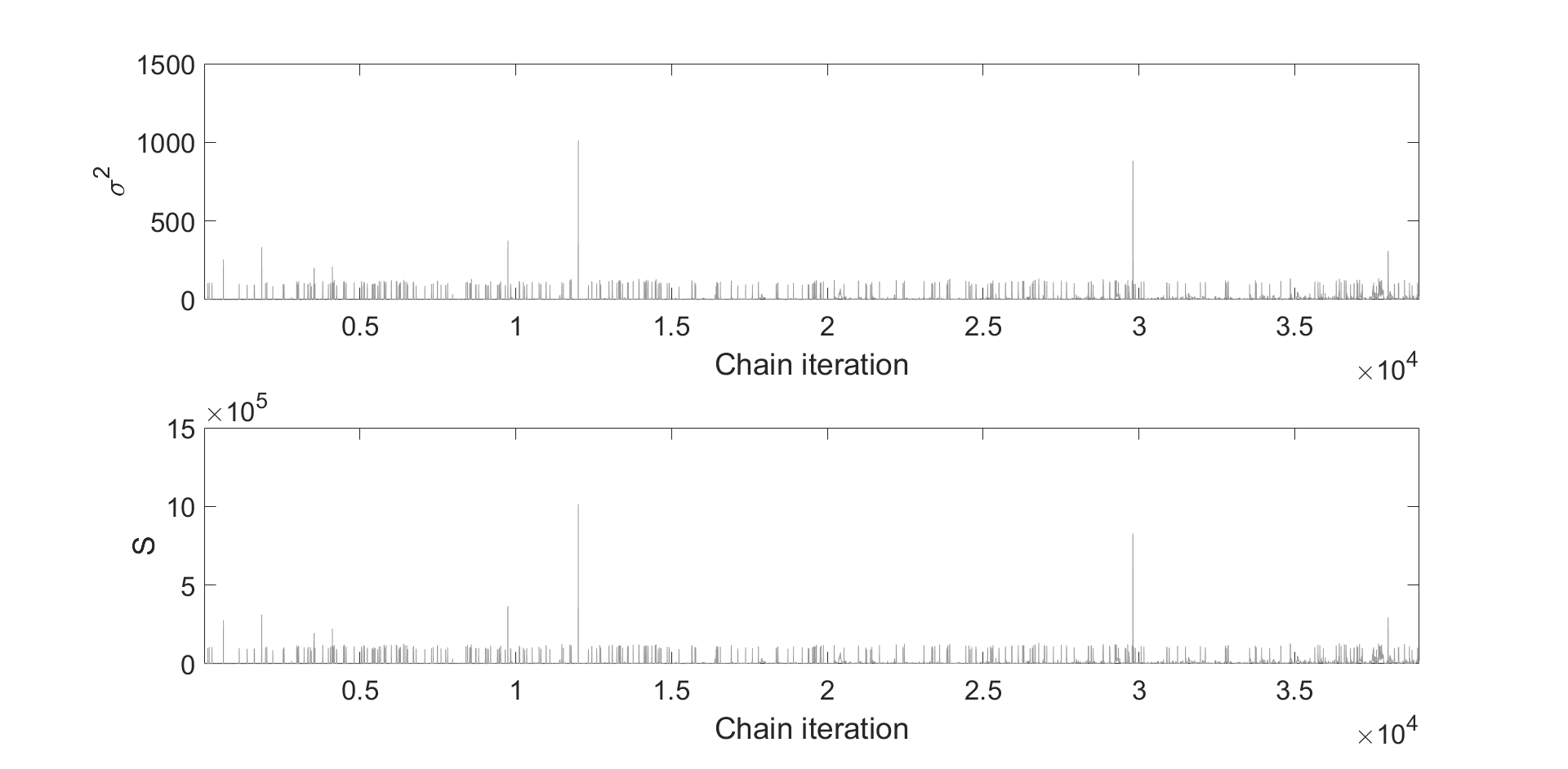}
\caption{Markov chains traceplots for the error variance $\sigma^2$, and the residual-sum-of-squares $S$ obtained obtained using the original Delayed Rejection Adaptive Metropolis algorithm in \cite{UncertaintySmith} without the correction term discussed in this paper, for the 5D model corresponding to a control mouse.}
\label{fig:20}
\end{figure}

Let us now illustrate (Figures \ref{fig:19} and \ref{fig:20}) what happens if we use the DRAM algorithm in \citep{UncertaintySmith}, that does not work in the transformed parameter space when calculating the acceptance probabilities, in particular the $2^{\textrm{nd}}$ stage acceptance probability.

When looking at Figure \ref{fig:20}, it is readily apparent that several proposals with high S are being falsely accepted. Figure \ref{fig:15} for the correct DRAM algorithm does not display such spikes, and the traceplots in Figure \ref{fig:14} are well-behaved. We falsely accept points because the ratio of the proposal distributions in equation (\ref{eq:7}) becomes very large, since it is calculated in the unscaled parameter space, and the starting values (optimum values) are very close to the boundaries or to the invalid paramater domain (see Figure \ref{fig:3}). This implies that many of the proposed points will either lie outside boundaries or inside the invalid domain, making $f$, the parameter with dominating magnitude, and S become high. When this ratio becomes large, the proposed point will be falsely accepted, as the acceptance rate is the minimum between 1 and the $\log$ of a high number, i.e. the acceptance probability will be 1. This would however not happen if the optimum value was somewhere far from the boundaries or the invalid domain, because the algorithm would explore a region of fairly similar values. This is tested for the case of the control mouse, model 4D, and the findings (not shown here) back up our hypothesis.   

\subsection{MCMC Convergence Diagnostics}
It is now of interest to perform convergence diagnostics on the DRAM chains for every mouse, for the two models. To do so, we employ the Geweke test, the Brooks Gelman Rubin test, the Integrated Autocorrelated Time (IACT), and the Effective Sample Size (ESS) tests and the results are summarised in Table \ref{table:2}.

The Geweke test reveals a high p-value (0.99), well above 0.05 for all 4 scenarios, suggesting that there is no evidence of a difference between the mean of the first $10\%$ samples and last $50\%$ samples in the chain. The MPSRF is below 1.1 for all 4 cases, the highest one being 1.03 and the lowest 1.0004. The 4D model for both mice provides a lower MPSRF $(< 1.01)$ than the 5D model. Nevertheless, the within and between chain covariance matrices are similar for both models. The estimates for IACT and ESS suggest that there is some correlation still left between the samples, in particular for the 5D model, where we have seen that there is a very strong correlation between $f$ and $\xi$. This strong correlation delays the algorithm from sampling independent draws. For the 4D model, for both mice, we notice that on average, one independent sample is drawn every 13 samples. Therefore, the ESS is somewhere in the range [1200, 1800] out of a total number of 20000 iterations. For the 5D model, there is a more pronounced correlation effect, with an average IACT of 27 and ESS ranging between [700, 1500]. While the Geweke test and the MPSRF do not indicate non-convergence of the chains, the IACT values (slightly too high) and ESS values (slightly too low) indicate that the DRAM algorithm should be run for longer in order to decrease the correlation between samples. 

When analysing Table \ref{table:3} which displays the MCMC convergence diagnostics for the AM algorithm, we notice that the AM algorithm produces even more correlated samples, the IACT being consistently higher than that for the DRAM algorithm. 

\subsection{Model Selection}
We further consider model selection, since we have two competing models for the two mice: a 4D model (without tapering factor, $\xi$ in), and a 5D model (with $\xi$ in). For the purpose of model selection, we show the performance of 4 criteria: AICc, BIC, DIC and WAIC.
We have seen in the previous section that there is still some correlation left between the chain samples. We therefore only consider a subset of 1000 samples of the total number of samples when computing WAIC.
Table \ref{table:4} summarises our model selection scores.

The results indicate that the scores are similar and fairly consistent across different criteria. For the control mouse, all the scores are lower  for the 5D model, compared to the 4D model, which implies that in the case of the control mouse, adding the tapering factor, $\xi$ makes the model more appropriate for the data, and so, it is worth including $\xi$ in the model. On the other hand, for the hypoxic mouse, the scores are very similar across the two models, WAIC favours the 4D model, while AICc, BIC and DIC favour the 5D model. Considering the very low difference in S between the 4D and 5D models (0.56), and the fact that there is low difference between the scores, then we choose the 4D model to be used in the case of the hypoxic mouse. Hence, there is evidence that the tapering factor is not significant for the hypoxic mouse, but the opposite is true for the control mouse. A possible explanation for this could be that in hypoxia, the vessel walls become stiff, and so there will not be a pronounced decrease in the radii along the vessels' length.

\section{Discussion}
In this paper we have applied the Delayed Rejection Adaptive Metropolis algorithm and its variations (Adaptive Metropolis and Delayed Rejection) to measure uncertainty of the parameter estimates in a partial differential equations system of the pulmonary circulation. The method, which builds on and further improves existing literature by allowing for novel parameter scaling due to different magnitudes of the parameters has been applied on real data, coming from a healthy and a diseased mouse.
Making a few restrictive assumptions about the parameters has allowed us to decrease the parameter dimension considerably (from a 55D problem -- $21 f, 11 r_1$, $11 r_2, 11 c, 1 \xi$ parameters, down to a 5D problem -- $1f, 1 r_1, 1 r_2, 1c, 1 \xi$). Further studies should be done to check these assumptions. Some preliminary exploration of non-constant Windkessel adjustments, $r_1, r_2, c$ and stiffness $f$ on a smaller network has indicated that fitting different sets of parameters for every vessel would result in data overfitting. Even if the goodness of fit has improved, the model complexity has increased significantly. Model selection criteria have revealed that a very low error variance would be needed to favour use of the more complex model. Since our error variance is believed to be higher, we have rejected the complex model. The parameters had different orders of magnitude, which motivated us to improve existing DRAM code from the literature in order to assess uncertainty in a system of partial differential equations, which is the pulmonary circulation. To make the computations feasible, we have started the MCMC algorithm from optimised parameter values (a nonlinear constraint optimisation scheme was used). Our findings indicate that DRAM is preferred over DR and AM. The DR algorithm alone is very slow in converging, and the chains appear far from converging even after 10000 iterations. The AM algorithm retains an acceptance rate of about $20\%$, which appears to be too low, and there is correlation between the samples even after 36000 iterations. The DRAM algorithm appears to be the best, with an acceptance rate of about $40\%$, and fairly good mixing. While the Geweke test and the Multivariate Scale Reduction Factor test did not indicate non-convergence, the Integrated Autocorrelation Time appeared to be slightly too high after 20000 iterations. The strong dependence between some of the parameters makes the algorithm slow in generating more independent samples, and should therefore be run for longer. Another focus of this paper was to perform model selection to identify whether one of the parameters should be included in the model. The AICc, BIC, DIC and WAIC criterion consistently chose the more complex model for the control mouse. For the hypoxic mouse, the scores were inconsistent, but since the difference between them was low, and the most reliable score, WAIC, favoured the simpler model, we preferred this model.   
All these results are based on a mathematical model that looks more appropriate for the hypoxic mouse than for the control mouse. Possible causes for the slight model mismatch are: (i) the simplicity of the model specifying the elastic behaviour of the blood vessels and/or the boundary conditions, (ii) uncertainty of the geometry measurements which are not specific to a given mouse, (iii) a combination of (i) and (ii). This slight model mis-specification has the undesired effect of breaking the i.i.d. error Normality assumption. Therefore, to account for the correlated structure in the errors, we aim to use a Markov model that will also model the correlation structure, and not only the parameters and the error variance. In addition, future work will include implementing faster methods to infer the relevant parameters. This includes the Hamiltonian Monte Carlo algorithm \citep{HMC}, which suppresses the random behaviour of the algorithm by using a momentum variable that leads the direction in which the proposals are made. To speed up the simulations, we will run this algorithm on an objective function emulated using Gaussian Processes \citep{GPs}.
Another alternative to learn the unobserved parameters is by employing the Ensemble Kalman filter, following the idea in \cite{EnKF}.

\section*{acknowledgements}
This work is part of the research programme of the Centre for multiscale soft tissue mechanics with application to heart \& cancer (SofTMech), funded by the Engineering and Physical Sciences Research Council (EPSRC) of the UK, grant reference number EP/N014642/1. Olufsen, Haider and Qureshi were supported by National Science Foundation (NSF-DMS \# 1615820). Data were made available by N. Chesler, Department of Biomedical Engineering, University of Wisconsin, Madison.

\clearpage
\bibliography{references}

\clearpage
\begin{table}
\begin{center} 
\begin{tabular}{| p{1.1cm} | p{1cm} | p{1cm} | p{1cm} | p{1cm} | p{1cm} | p{1cm} | p{1cm} |}
  \headrow
  \thead{Type} & \thead{$f$} & \thead{$r_1$} & \thead{$r_2$} & \thead{$c$} & \thead{$\xi$} & \thead{$\sigma^2$} & \thead{$S$} \\
 \hline
 \hiderowcolors
 Control 4D & 100374 (391) & 1.69 (0.01) & 0.17 (0.005) & -1.30 (0.04) & - & 0.12 (0.005) & 127 \\ 
 \hline
 Control 5D & 85471 (658) & 1.96 (0.006) & 0.25 (0.01) & -2.03 (0.07) & 0.26 (0.01) & 0.10 (0.004) & 105 \\ 
 \hline
 Hypoxic 4D & 243653 (1180) & 0.52 (0.01) & 0.13 (0.003) & -0.24 (0.01) & - & 0.08 (0.003) & 80.56 \\ 
 \hline
Hypoxic 5D & 231627 (3720) & 0.57 (0.02) & 0.14 (0.003) & -0.23 (0.01) & 0.07 (0.02) & 0.08 (0.003) & 80 \\ 
 \hline
\end{tabular}
\caption{Parameter ($f, r_1, r_2, c, \xi$) estimates and $\sigma^2$ estimate with their standard deviations inside brackets and S (residual-sum-of squares) for every type of mouse (control and hypoxic) and model (4D and 5D) obtained from running the Adaptive Metropolis algorithm. These parameters are defined in the Simulations section.}
\label{table:1}
\end{center}
\end{table}

\clearpage
\begin{table}
\begin{center} 
\begin{tabular}{| p{1.5cm} | p{1.5cm} | p{1.5cm} | p{2.5cm} | p{3.5cm} |}
  \hline
  \headrow
  \thead{Type} & \thead{Geweke test (p-values)} & \thead{MPSRF} & \thead{IACT} & \thead{ESS (out of 20000)} \\
 \hline
 \hiderowcolors
 Control 4D & all $>$ 0.05 & 1.004 & (11,12,11,14) & (1443,1362,1424,1180) \\ 
 \hline
 Control 5D & all $>$ 0.05 & 1.01 & (22,31,28,21,22) & (912,639,702,930,927) \\ 
 \hline
 Hypoxic 4D & all $>$ 0.05 & 1.006 & (14,15,14,11) & (1412,1333,1465,1807) \\ 
 \hline
Hypoxic 5D & all $>$ 0.05 & 1.03 & (23,22,13,17,27) & (862,897,1530,1144,732) \\ 
 \hline
\end{tabular}
\caption{MCMC Convergence Diagnostics: Geweke test, Multivariate Potential Scale Reduction Factor (MPSRF), Integrated Autocorrelation Time (IACT) for each of the parameters, Effective Sample Size (ESS) for each of the parameters, corresponding to the results obtained using the Delayed Rejection Adaptive Metropolis algorithm for the two mice (control, hypoxic) and the two models (4D containing 4 parameters and 5D containing 5 parameters).}
\label{table:2}
\end{center}
\end{table}

\clearpage
\begin{table}
\begin{center} 
\begin{tabular}{| p{1.5cm} | p{1.5cm} | p{1.5cm} | p{2.5cm} | p{3.5cm} |}
  \hline
  \headrow
  \thead{Type} & \thead{Geweke test (p-values)} & \thead{MPSRF} & \thead{IACT} & \thead{ESS (out of 36000)} \\
 \hline
 \hiderowcolors
 Control 4D & all $>$ 0.05 & 1.01 & (22,22,20,22) & (1630,1671,1811,1625) \\ 
 \hline
 Control 5D & all $>$ 0.05 & 1.04 & (39,47,49,27,43) & (918,767,731,1317,837) \\ 
 \hline
 Hypoxic 4D & all $>$ 0.05 & 1.004 & (17,18,18,18) & (2123,2030,2056,1990) \\ 
 \hline
Hypoxic 5D & all $>$ 0.05 & 1.06 & (49,37,19,26,58) & (730,965,1899,1375,622) \\ 
 \hline
\end{tabular}
\caption{MCMC Convergence Diagnostics: Geweke test, Multivariate Potential Scale Reduction Factor (MPSRF), Integrated Autocorrelation Time (IACT) for each of the parameters, Effective Sample Size (ESS) for each of the parameters, corresponding to the results obtained using the Adaptive Metropolis algorithm for the two mice (control, hypoxic) and the two models (4D containing 4 parameters and 5D containing 5 parameters).}
\label{table:3}
\end{center}
\end{table}

\clearpage
\begin{table}
\begin{center}
\begin{tabular}{| p{1.5cm} | p{1.5cm} | p{1.5cm} | p{1.5cm} | p{1.5cm} |}
  \hline
  \headrow
\thead{Type} & \thead{AICc (AIC)} & \thead{BIC} & \thead{DIC} & \thead{WAIC}\\
 \hline
 Control 4D & 772 & 792 & 772 & 774 \\
 \hiderowcolors
 \hline
 Control 5D & 579 & 603 & 578 & 580 \\ 
 \hline
 Hypoxic 4D & 307 & 326 & 306 & 304 \\ 
 \hline
 Hypoxic 5D & 301 & 325 & 300 & 310 \\ 
 \hline
\end{tabular}
\caption{Model selection scores (AICc, BIC, DIC, WAIC) for the 4D and 5D models for the control and the hypoxic mouse. AICc is very similar to AIC (up to 4 dp), since the sample size is large (1024 data points). For the calculation of AICc and BIC, we have estimated $\hat{\sigma}^2 = \frac{S}{n-d}$, for DIC, $\hat{\sigma}^2$ is the posterior mean of the $\sigma^2$ samples, and for WAIC, every $\sigma^2$ sample in the chain is used.}
\label{table:4}
\end{center}
\end{table}

\clearpage
\begin{appendices}

\begin{figure}[h!]
\centering
\includegraphics[scale=0.25]{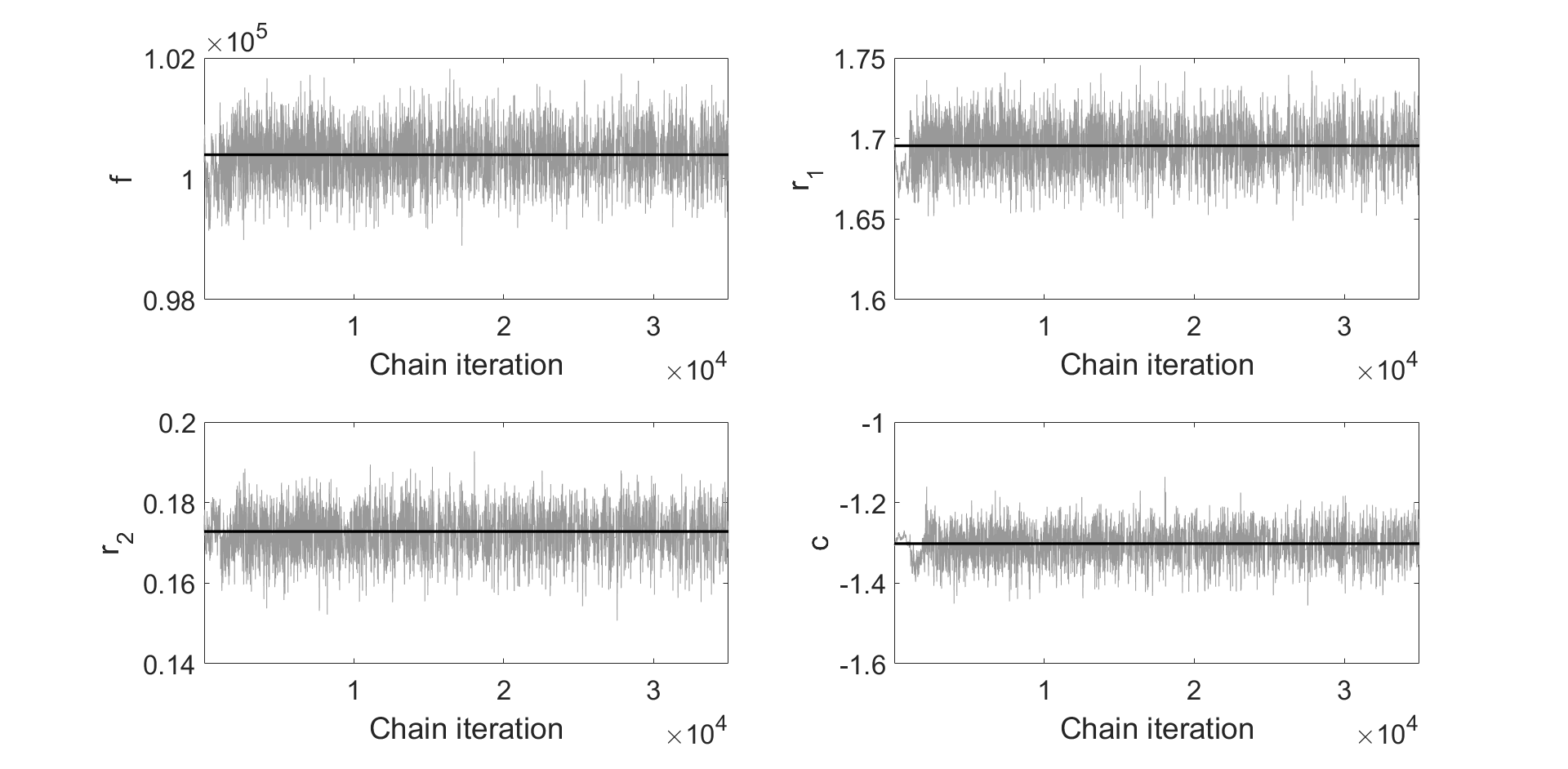}
\caption{Markov chains traceplots for the parameters obtained using the Adaptive Metropolis algorithm for the 4D model corresponding to a control mouse. Starting values for the algorithm are the optimised values and are superimposed in black horizontal lines. Acceptance rate is $20\%$.}
\label{fig:10}
\end{figure}

\begin{figure}[h!]
\centering
\includegraphics[scale=0.25]{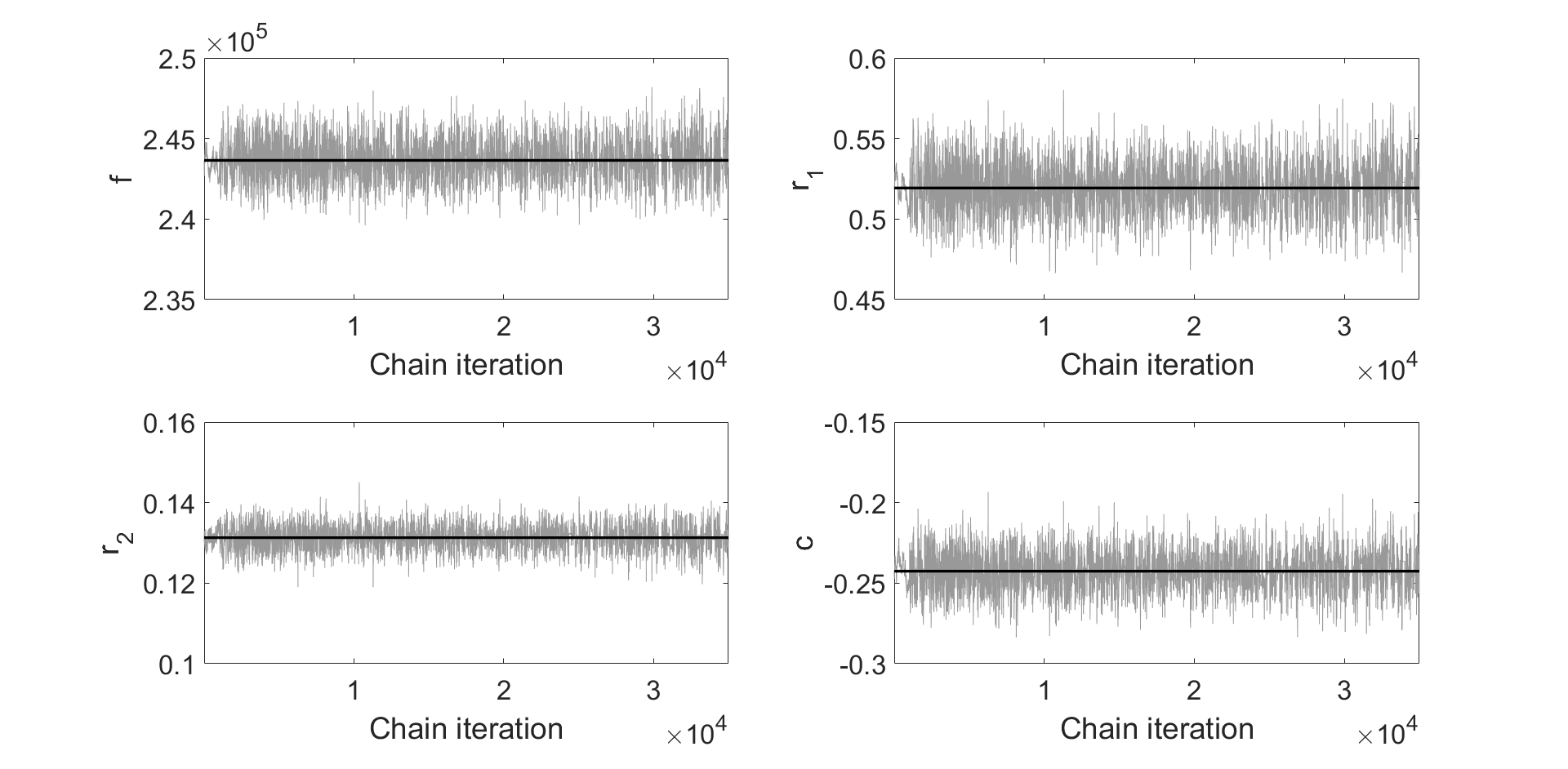}
\caption{Markov chains traceplots for the parameters (defined in the Simulations section) obtained using the Adaptive Metropolis algorithm for the 4D model corresponding to a hypoxic mouse. Starting values for the algorithm are the optimised values and are superimposed in black horizontal lines. Acceptance rate is $19\%$.}
\label{fig:12}
\end{figure}

\begin{figure}[h!]
\centering
\includegraphics[scale=0.25]{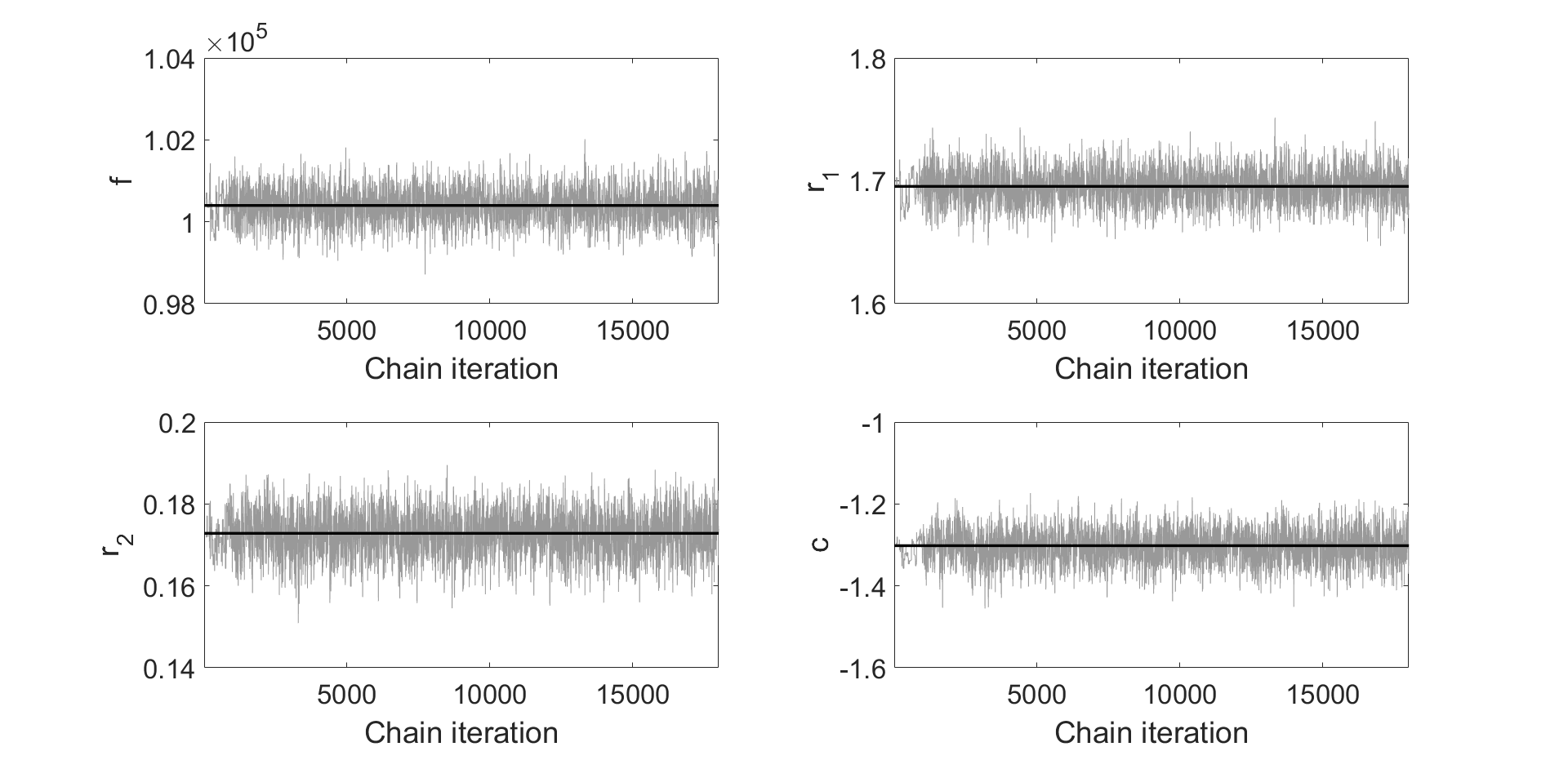}
\caption{Markov chains traceplots for the parameters (defined in the Simulations section) obtained using the Delayed Rejection Adaptive Metropolis algorithm for the 4D model corresponding to a control mouse. Starting values for the algorithm are the optimised values and are superimposed in black horizontal lines. Acceptance rate is $48\%$.}
\label{fig:16}
\end{figure}

\begin{figure}[h!]
\centering
\includegraphics[scale=0.25]{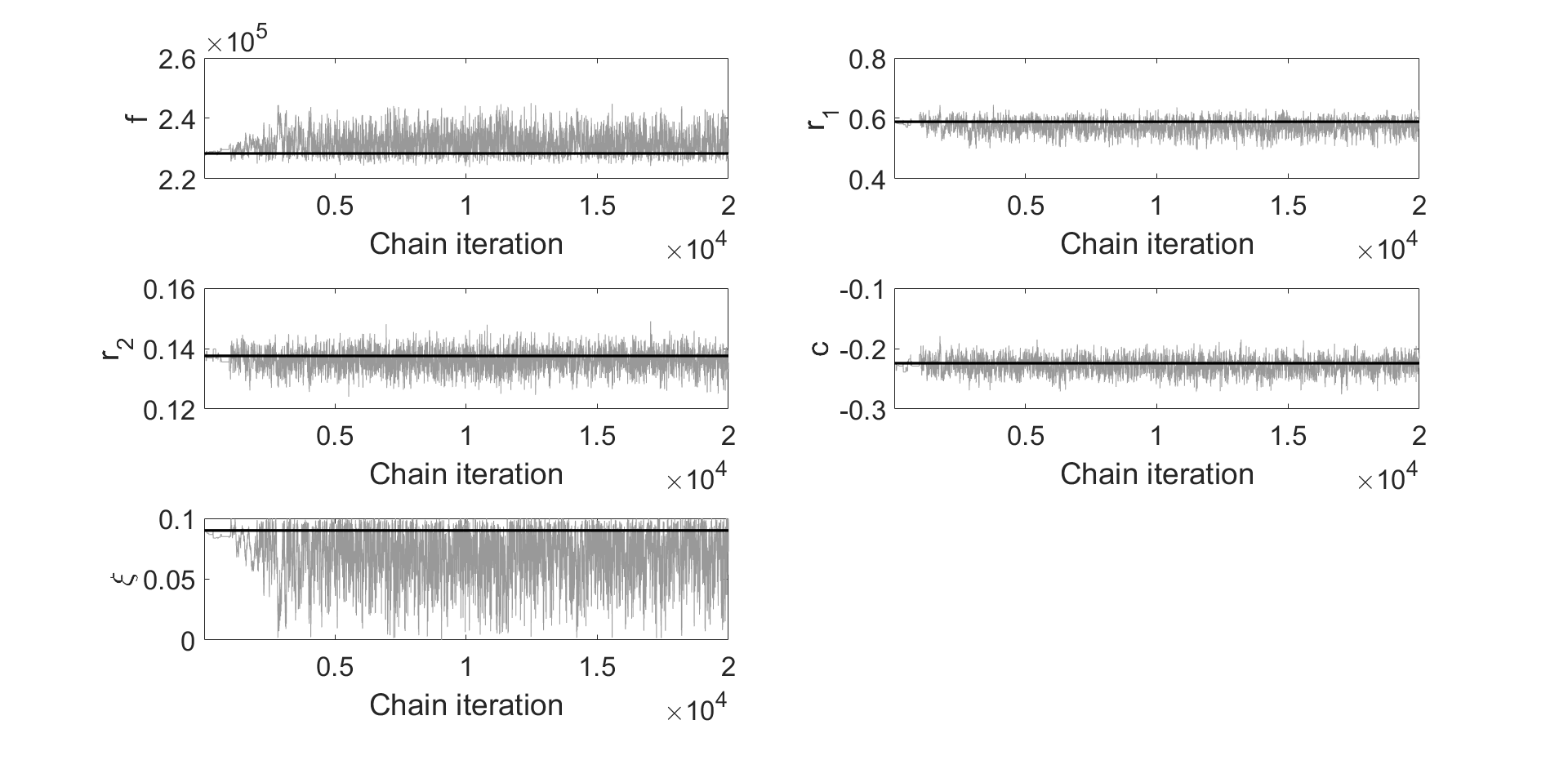}
\caption{Markov chains traceplots for the parameters (defined in the Simulations section) obtained using the Delayed Rejection Adaptive Metropolis algorithm for the 5D model corresponding to a hypoxic mouse. Starting values for the algorithm are the optimised values and are superimposed in black horizontal lines. Acceptance rate is $45\%$.}
\label{fig:17}
\end{figure}

\begin{figure}[tb]
\centering
\includegraphics[scale=0.25]{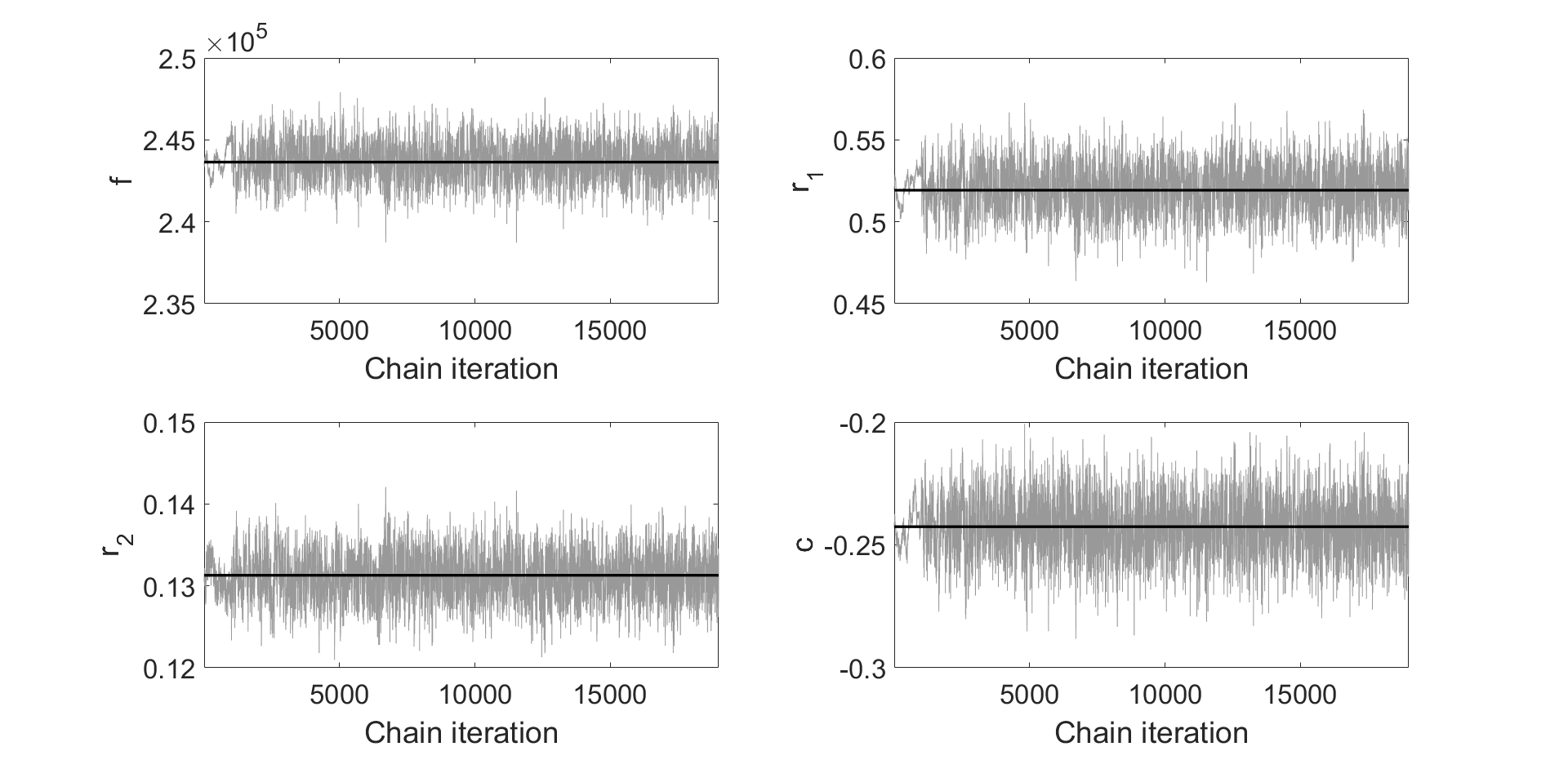}
\caption{Markov chains traceplots for the parameters (defined in the Simulations section) obtained using the Delayed Rejection Adaptive Metropolis algorithm for the 4D model corresponding to a hypoxic mouse. Starting values for the algorithm are the optimised values and are superimposed in black horizontal lines. Acceptance rate is $46\%$.}
\label{fig:18}
\end{figure}

\end{appendices}

\end{document}